\documentclass[journal]{IEEEtran}
\usepackage{amsmath,amsfonts}
\usepackage{algorithmic}
\usepackage{algorithm}
\usepackage{array}
\usepackage{textcomp}
\usepackage{stfloats}
\usepackage[hyphens]{url}
\usepackage{verbatim}
\usepackage{subfigure} %
\usepackage{amssymb} %
\usepackage{multirow} %
\usepackage{makecell} %
\usepackage{ragged2e} %
\usepackage{color} %
\usepackage[switch]{lineno}
\usepackage{graphicx}
\usepackage{cite}
\begin{document}
\bibliographystyle{IEEEtran}

\title{Adaptive Convolution for CNN-based Speech Enhancement Models}

\author{Dahan Wang, Xiaobin Rong, Shiruo Sun, Yuxiang Hu, Changbao Zhu, and Jing Lu
\thanks{Published in IEEE Transactions on Audio, Speech and Language Processing. DOI: 10.1109/TASLPRO.2025.3623897. This work was supported by the National Natural
Science Foundation of China (Grant No. 12274221) and the AI \& AI for Science Project of Nanjing University. The associate editor coordinating the review of this article and approving it for publication was Dr. Marc Delcroix. (Corresponding author: Jing Lu.)}
\thanks{Dahan Wang, Xiaobin Rong, and Jing Lu are with the Key Laboratory of Modern Acoustics, Institute of Acoustics, Nanjing University, Nanjing 210093, China, and also with the NJU-Horizon Intelligent Audio Lab, Horizon Robotics, Beijing 100094, China (e-mail: dahan.wang@smail.nju.edu.cn; xiaobin.rong@smail.nju.edu.cn; lujing@nju.edu.cn). Shiruo Sun, Yuxiang Hu, and Changbao Zhu are with the NJU-Horizon Intelligent Audio Lab, Horizon Robotics, Beijing 100094, China (e-mail: shiruo.sun@gua.com; yuxiang.hu@gua.com; changbao.zhu@gua.com).}}

\markboth{IEEE TRANSACTIONS ON AUDIO, SPEECH AND LANGUAGE PROCESSING, VOL. 33, 2025}%
{Shell \MakeLowercase{\textit{et al.}}: A Sample Article Using IEEEtran.cls for IEEE Journals}

\IEEEpubid{
\begin{minipage}{1.1\textwidth}
\vspace{20pt}\centering
2998-4173~\copyright~2025 IEEE. All rights reserved, including rights for text and data mining, and training of artificial intelligence and similar technologies.\\ Personal use is permitted, but republication/redistribution requires IEEE permission. See \url{https://www.ieee.org/publications/rights/index.html} for more information.
\end{minipage}}

\maketitle

\begin{abstract}
Deep learning-based speech enhancement methods have significantly improved speech quality and intelligibility. Convolutional neural networks (CNNs) have been proven to be essential components of many high-performance models. In this paper, we introduce adaptive convolution, an efficient and versatile convolutional module that enhances the model’s capability to adaptively represent speech signals. Adaptive convolution performs frame-wise causal dynamic convolution, generating time-varying kernels for each frame by assembling multiple parallel candidate kernels. A lightweight attention mechanism is proposed for adaptive convolution, leveraging both current and historical information to assign adaptive weights to each candidate kernel. This enables the convolution operation to adapt to frame-level speech spectral features, leading to more efficient extraction and reconstruction. We integrate adaptive convolution into various CNN-based models, highlighting its generalizability. Experimental results demonstrate that adaptive convolution significantly improves the performance with negligible increases in computational complexity, especially for lightweight models. Moreover, we present an intuitive analysis revealing a strong correlation between kernel selection and signal characteristics. Furthermore, we propose the adaptive convolutional recurrent network (AdaptCRN), an ultra-lightweight model that incorporates adaptive convolution and an efficient encoder-decoder design, achieving superior performance compared to models with similar or even higher computational costs.
\end{abstract}


\section{Introduction \label{sec1}}
Speech enhancement (SE) aims to improve the quality and intelligibility of speech signals contaminated by background noise. It serves as a critical front-end processing task for various applications, including mobile communications, teleconferencing, earphones, hearing aids, and automatic speech recognition systems \cite{benesty2006speech}. Traditional SE methods \cite{cohen2001speech,wang2024incorporation}, primarily based on signal processing techniques, usually struggle in non-stationary noise conditions. Over the past decade, deep learning-based methods have demonstrated remarkable success in addressing diverse and complex acoustic scenarios by leveraging deep neural networks (DNNs) to recover clean speech in an end-to-end manner. However, the superior performance of DNN-based methods is typically accompanied by high computational complexity and large model sizes. Most state-of-the-art (SOTA) models \cite{hao2021fullsubnet,wang2023tf} require considerable computational resources, often involving billions of multiply-accumulate (MAC) operations per second, which limits their deployment on resource-constrained edge devices. While many lightweight SE models suitable for real-time applications \cite{choi2021real,rong2024gtcrn} have been proposed in recent years, they may encounter performance degradation in challenging acoustic environments.

Convolutional neural networks (CNNs) have been widely utilized in SE tasks, particularly as encoder-decoder architectures for extracting and reconstructing speech features. Convolutional recurrent network (CRN) \cite{tan2018convolutional} operates in the time-frequency domain, employing a convolutional encoder-decoder and recurrent neural network (RNN) to capture local spectrogram patterns and model the dependence between consecutive frames. Deep complex CRN (DCCRN) \cite{hu2020dccrn} enhances the performance of CRN by incorporating complex-valued operations to better handle the phase information of speech signals. Dual-path CRN (DPCRN) \cite{le2021dpcrn} extends the CRN architecture by integrating dual-path RNN (DPRNN) \cite{luo2020dual}, improving the modeling of spectral patterns within individual frames. DeepFilterNet \cite{schroter2022deepfilternet,schroter2022deepfilternet2} achieves two-stage enhancement of envelope and periodic components through two encoder-decoders combined with deep filtering technique.
\IEEEpubidadjcol

Recently, the development of several ultra-lightweight models with computational complexity below 100 MMACs has demonstrated the potential of lightweight CRN architectures in effectively addressing the challenges posed by complex acoustic conditions. Grouped temporal CRN (GTCRN) \cite{rong2024gtcrn} efficiently simplifies DPCRN by incorporating grouped strategies and introduces subband feature extraction (SFE) and temporal recurrent attention (TRA) modules to boost performance. Another network proposed in \cite{yang2024fspen}, FSPEN, utilizes fullband and subband encoder-decoder alongside an inter-frame path extension approach to improve feature extraction and modeling efficiency with a low computational cost. ULCNet \cite{shetu2024ultra} adopts a two-stage processing framework, incorporating channel-wise feature reorientation and a modified power-law compression technique to achieve both reduced computational costs and enhanced speech quality. Moreover, LiSenNet \cite{yan2024lisennet} leverages various modules and techniques, including subband convolution, convolutional gated linear unit, noise detection, and phase refinement, to optimize its lightweight design.

The design of convolutional blocks is a critical aspect of CRN-based model development. For lightweight SE models, convolutional encoder-decoder designs often emphasize leveraging fullband and subband features of speech spectrums \cite{yang2024fspen,yan2024lisennet}, while paying less attention to the general architectural design of convolutional blocks. In contrast, the design of CNNs has been extensively explored in the field of computer vision (CV). ResNet \cite{he2016deep} introduces a residual block specifically designed for deeper networks, providing a refined convolutional block structure. MobileNetV2 \cite{sandler2018mobilenetv2} proposes the inverted residual block, reducing parameters and computational complexity with minimal accuracy loss. ConvNeXt \cite{liu2022convnet} further ``modernized'' the ResNet architecture, enabling CNNs to rival or even outperform Transformer-based models in certain tasks. Additionally, recent research \cite{ma2024rewrite} highlights the potential of the ``star operation'' (element-wise multiplication), which underpins the design of StarNet, achieving high performance with low latency.

The most straightforward way to increase a model's capacity is to increase its width and depth, i.e., by adding more channels and layers. However, this approach inevitably results in a significant increase in computational complexity. Dynamic convolution (DyConv) \cite{chen2020dynamic} is an effective method to address this challenge. It enhances the representation capability of models by dynamically aggregating multiple parallel candidate convolution kernels based on input-dependent attentions, without increasing the network’s depth or width. Due to the utilization of lightweight attention mechanism, DyConv introduces minimal computational overhead at the expense of an increased number of parameters. ParameterNet \cite{han2024parameternet} leverages this characteristic to improve model's capacity, addressing the challenge that models with low computational cost cannot benefit from large-scale pretraining in vision and language tasks. Omni-dimensional dynamic convolution (ODConv) \cite{li2022odconv} improves the attention mechanism by applying complementary attention across spatial, channel, filter, and candidate kernel dimensions of the kernel space, achieving comparable performance to DyConv with fewer candidate kernels. KernelWarehouse \cite{li2024kernelwarehouse} generalizes dynamic convolution further, achieving an effective trade-off between parameter efficiency and representation capability.

Dynamic convolution is initially designed for CV tasks, where it aggregates information from an entire image to compute kernel attention weights and applies a fixed kernel configuration at the image level rather than the pixel level \cite{chen2020dynamic}. A related approach, adaptive convolution, has also been proposed in the CV field, generating pixel-level independent kernels for each spatial location in the feature map \cite{su2019pixel, xu2020squeezesegv3, xu2021paconv}. However, these methods are not directly suitable for causal SE. The former introduces strong non-causality, while the latter typically relies on complex architectures or additional information for kernel computation, leading to excessive computational cost. Moreover, for highly non-stationary noisy speech, an utterance-level kernel configuration may be suboptimal for different signal segments or frames. One intuitive understanding is that the optimal convolution kernels for speech frames and noise frames are likely to differ significantly. This suggests that, for real-time SE, frame-level causal kernel adjustment could better adapt to time-varying signal characteristics while maintaining low computational overhead. This idea has been briefly explored in dynamic gate convolutional networks (DGCN) \cite{chen2022dgcn}, where the global pooling in the squeeze-and-excitation mechanism of dynamic convolution is replaced with pooling along the frequency dimension within each frame to produce time-varying kernels. However, in DGCN, this mechanism is applied only to a subset of convolutional modules, without further refinement, systematic exploration, or extensive experimentation. In contrast to DGCN, our work thoroughly explores, refines, and extends the attention mechanism of this approach, demonstrating its generalizability across diverse baseline models. In particular, we emphasize its critical role in improving the performance of lightweight networks and simultaneously propose an ultra-lightweight model for real-time SE. Furthermore, we reveal a key connection between dynamic kernel adjustment and speech spectral characteristics. The main contributions of this paper are summarized in the following paragraphs.

In this paper, we propose adaptive convolution, an efficient and versatile convolution mechanism that performs frame-wise causal dynamic convolution through an enhanced attention mechanism explicitly leveraging historical information. Adaptive convolution generates time-varying kernels for each frame rather than relying on a fixed kernel configuration across different input features. Given that convolution kernels can be regarded as filters, the proposed method is analogous to adaptive filtering, as both approaches adjust filter coefficients in real time based on the statistical characteristics of the input signal. The adaptive convolution kernel is obtained by aggregating multiple parallel candidate kernels through frame-level attention weights derived from input features. This enables the convolution operation to adapt to frame-level speech spectral features, facilitating more efficient feature extraction and reconstruction. The attention mechanism plays a critical role in ensuring the effectiveness of adaptive convolution. We explore several channel modeling approaches for kernel attention, including more lightweight single-frame and multi-frame modeling, as well as higher-performance temporal modeling. Additionally, for convolutional blocks composed of multiple depthwise and pointwise convolution layers, we propose a multi-head mechanism to jointly generate kernel attention for these layers. This mechanism can also simultaneously generate optional input and output temporal channel attention maps, further enhancing the capability to model non-stationary speech signals.

In this paper, we highlight the effectiveness and generalizability of adaptive convolution across diverse model architectures. We conduct extensive experiments on several CNN-based models of diverse structures and scales, including DPCRN \cite{le2021dpcrn} of different sizes, DCCRN \cite{hu2020dccrn}, GTCRN \cite{rong2024gtcrn}, and LiSenNet \cite{yan2024lisennet}, by replacing vanilla convolutions with adaptive convolutions. Experimental results demonstrate that adaptive convolution is an efficient and versatile convolutional module that significantly enhances the performance of various models, particularly those with lower computational burdens. Although adaptive convolution increases the number of parameters, the resulting increment in computational complexity, which is more crucial for many applications, is limited. Ablation experiments also highlight the advantages of the proposed adaptive convolution over non-causal, utterance-level dynamic convolution, underscoring the importance of frame-wise kernel adjustment. Additionally, through a detailed analysis of the kernel attention weights, we clearly demonstrate that adaptive convolution can effectively assign appropriate candidate kernels to frames with varying spectral features.

To further illustrate the practical value of the proposed adaptive convolution, we propose an ultra-lightweight model, adaptive convolution recurrent network (AdaptCRN). Inspired by ConvNeXt \cite{liu2022convnet} and StarNet \cite{ma2024rewrite} blocks, we design a lightweight and efficient encoder-decoder architecture that integrates adaptive convolution. We also incorporate strategies such as grouped RNN \cite{rong2024gtcrn, gao2018efficient} and spectral compression to further enhance performance and reduce model scale. Experimental results show that AdaptCRN achieves competitive performance with only {41M} MACs per second and 135K parameters, outperforming several other SOTA lightweight models with similar or even significantly higher computational cost.

This paper is organized as follows: Section~\ref{sec2} reviews the dynamic convolution. Section~\ref{sec3} introduces the proposed adaptive convolution, detailing its attention mechanism and implementation. Section~\ref{sec4} presents the proposed ultra-lightweight model, AdaptCRN. Section~\ref{sec5} describes the experimental setup. Section~\ref{sec6} reports the experimental results, verifying the effectiveness and generalization of adaptive convolution, evaluating the performance of AdaptCRN, and demonstrating the relationship between the kernel attention of adaptive convolution and speech spectral features. Finally, Section~\ref{sec7} draws the conclusions of the paper.

\section{Related work \label{sec2}}
\subsection{Dynamic Convolution \label{sec2.1}}
DyConv \cite{chen2020dynamic} is an effective method to enhance the representation capability of models with negligible additional computational complexity. It employs input-dependent dynamic kernels instead of static counterparts, making the kernels nonlinear functions of the input features. The dynamic kernels are generated by aggregating a set of parallel candidate kernels with identical kernel size $K_\text{h} \times K_\text{w}$ and input/output channels  $C_{\text{in/out}}$. The kernel aggregation is guided by the attention weights and can be expressed as
\begin{equation}
    \mathbf{W} = \sum_{k=1}^{K} A_k \mathbf{W}_k,
    \label{eq.1}
\end{equation}
where $\mathbf{W}$ represents the resulting dynamic kernel, $K$ is the number of candidate kernels, $\mathbf{W}_k, k=1,2,\cdots,K$ denotes the $k$-th candidate kernel consisting of $C_{\text{out}}$ filters $\mathbf{W}_k^m \in \mathbb{R}^{K_\text{h} \times K_\text{w} \times C_{\text{in}}}, m=1,2,\cdots,C_{\text{out}}$, and $A_k \in \mathbb{R}$ is the attention weight assigned to $\mathbf{W}_k$. The attention weights are computed using a squeeze-and-excitation mechanism. Specifically, global spatial information is first compressed through global average pooling (GAP), followed by a fully connected (FC) layer, a rectified linear unit (ReLU) activation, another FC layer, and a softmax function. Due to the small size of the convolution kernels and the lightweight nature of the attention mechanism, DyConv is computationally efficient at the cost of an increased number of parameters. Dynamic convolution similarly handles bias, but for simplicity, this detail is not included in the formulas.

The softmax function ensures that the attention weights sum to one, as expressed by $\sum_{k=1}^{K} A_k = 1$. As pointed out in DyConv, this normalization helps constrain the kernel space, facilitating the learning of the attention model. During training, candidate kernels with low attention weights are often difficult to optimize. To address this, DyConv incorporates temperature annealing in the softmax function, encouraging a near-uniform attention distribution in the early training stages. Specifically, the features are scaled by a temperature coefficient, which varies during training, before applying the softmax function. A higher temperature produces a flatter distribution of attention weights, facilitating the learning of all kernels. As training progresses, the temperature is gradually reduced, allowing the model to assign more distinct attention weights and improving its ability to select the most relevant kernels.

\subsection{Omni-dimensional Dynamic Convolution \label{sec2.2}}
ODConv \cite{li2022odconv} introduces a multi-dimensional attention mechanism to enhance the performance of DyConv. It employs a parallel strategy to learn complementary attention weights for convolutional kernels across four dimensions of the kernel space: spatial, channel (input channel), filter (output channel), and candidate kernel dimensions. Specifically, in addition to kernel attention, ODConv assigns distinct attention weights, denoted as $\mathbf{A}_k^\text{s} \in \mathbb{R}^{K_\text{h} \times K_\text{w}}$, $\mathbf{A}_k^\text{c} \in \mathbb{R}^{C_{\text{in}}}$, and $\mathbf{A}_k^\text{f} \in \mathbb{R}^{C_{\text{out}}}$, to different spatial locations, input channels, and output channels of the $k$-th candidate kernel. It can be represented as
\begin{equation}
    \mathbf{W} = \sum_{k=1}^{K} A_k \left( \mathbf{A}_k^\text{s} * \mathbf{A}_k^\text{c} * \mathbf{A}_k^\text{f} * \mathbf{W}_k \right),
    \label{eq.2}
\end{equation}
where $*$ denotes element-wise multiplication, with implicit broadcasting of the attention weights to align with the shape of $\mathbf{W}_k$. The attention weights for each dimension are computed using a multi-head mechanism. Specifically, the final FC layer in the squeeze-and-excitation attention module described in Section~\ref{sec2.1} is extended to 4 parallel FC layers, with the additional 3 layers followed by sigmoid functions to generate the spatial, channel, and filter attention weights. For simple implementation, these weights can be shared across all candidate kernels, allowing the subscript $k$ of the corresponding attention descriptor in Eq.~\ref{eq.2} to be omitted. By leveraging this refined kernel adjustment mechanism, ODConv achieves comparable or superior performance to standard DyConv while requiring fewer convolutional kernels and significantly fewer extra parameters.

\section{Adaptive Convolution \label{sec3}}
Considering the time-varying nature of speech signals and the real-time SE requirements for causality and computational efficiency, we propose adaptive convolution, a frame-wise causal dynamic convolution mechanism, which enables the convolution operation to adapt to frame-level speech spectral features. Similar to adaptive filtering, adaptive convolution dynamically adjusts the convolutional kernel in real time based on the statistical characteristics of the input signal. As depicted in Fig.~\ref{fig1}(a), the adaptive kernel is generated by aggregating multiple parallel candidate kernels with frame-level attention weights derived from input features, represented as
\begin{equation}
    \mathbf{W}\left(t\right) = \sum_{k=1}^{K} A_k\left(t\right) \mathbf{W}_k,
    \label{eq.3}
\end{equation}
where $t$ denotes the frame index, emphasizing the time-varying nature of the kernel. The learnable candidate kernels, $\mathbf{W}_k, k=1,2,\cdots,K$, are end-to-end optimized and then fixed after training. The adaptive kernel is then applied in a causal convolution with the features of the corresponding frame to produce the output. For simplicity, we keep the biases of the convolution layers static rather than adaptive.

\subsection{Kernel Attention \label{sec3.1}}
\begin{figure}[!t]
    \centering
    \subfigure[]{\includegraphics[height=2.1in]{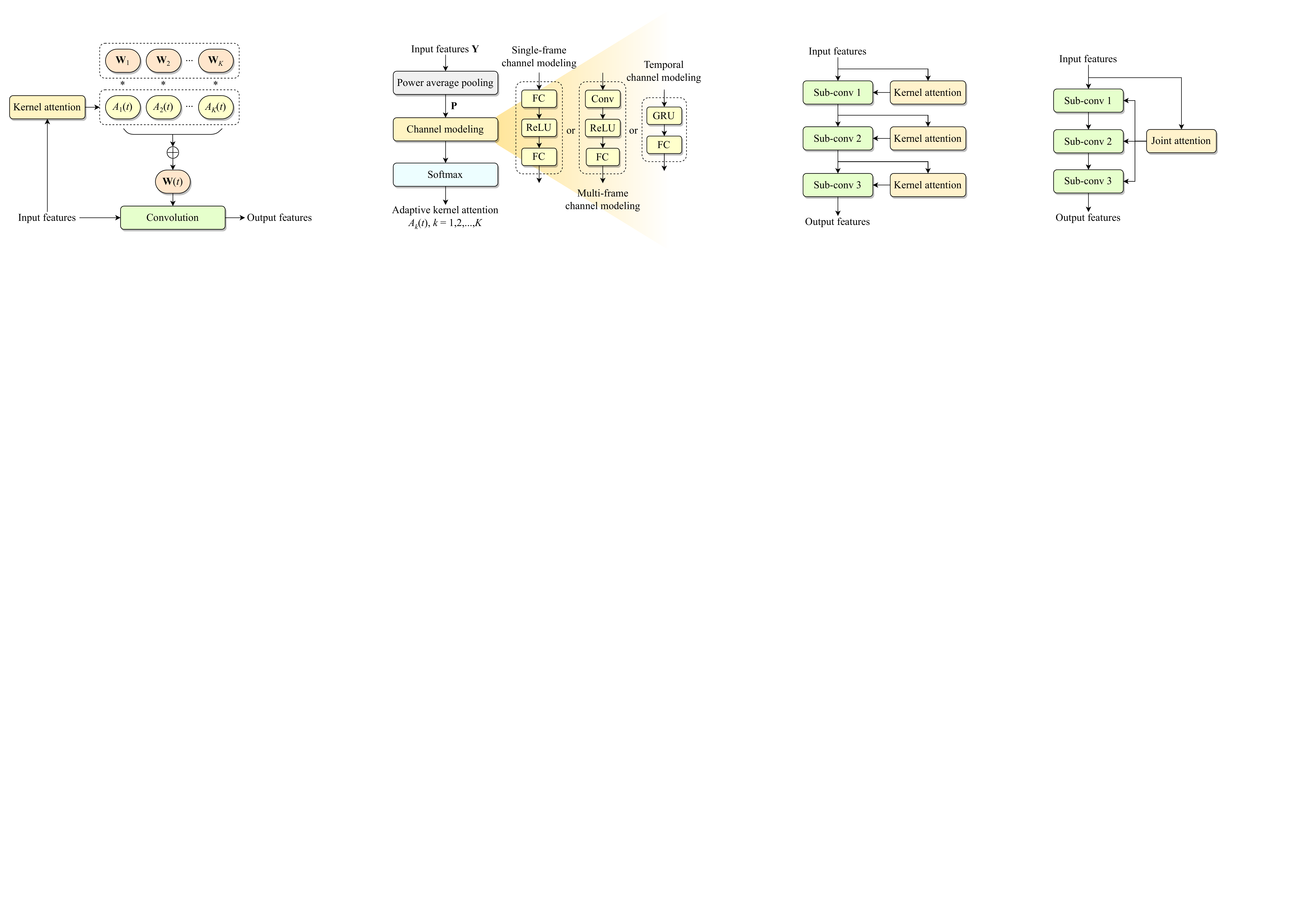}} \\
    \subfigure[]{\includegraphics[height=2.1in]{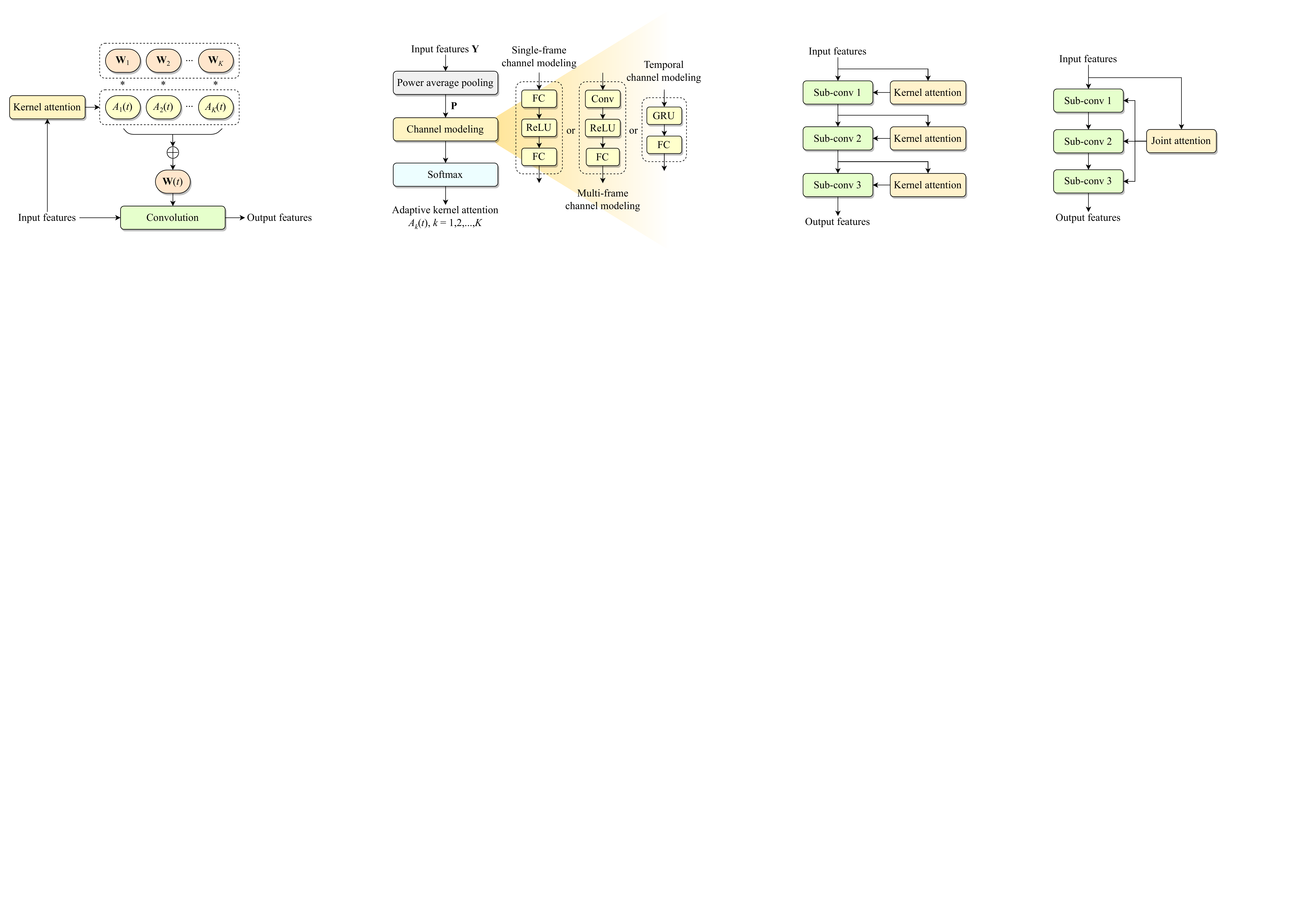}} \\
    \subfigure[]{\includegraphics[height=1.8in]{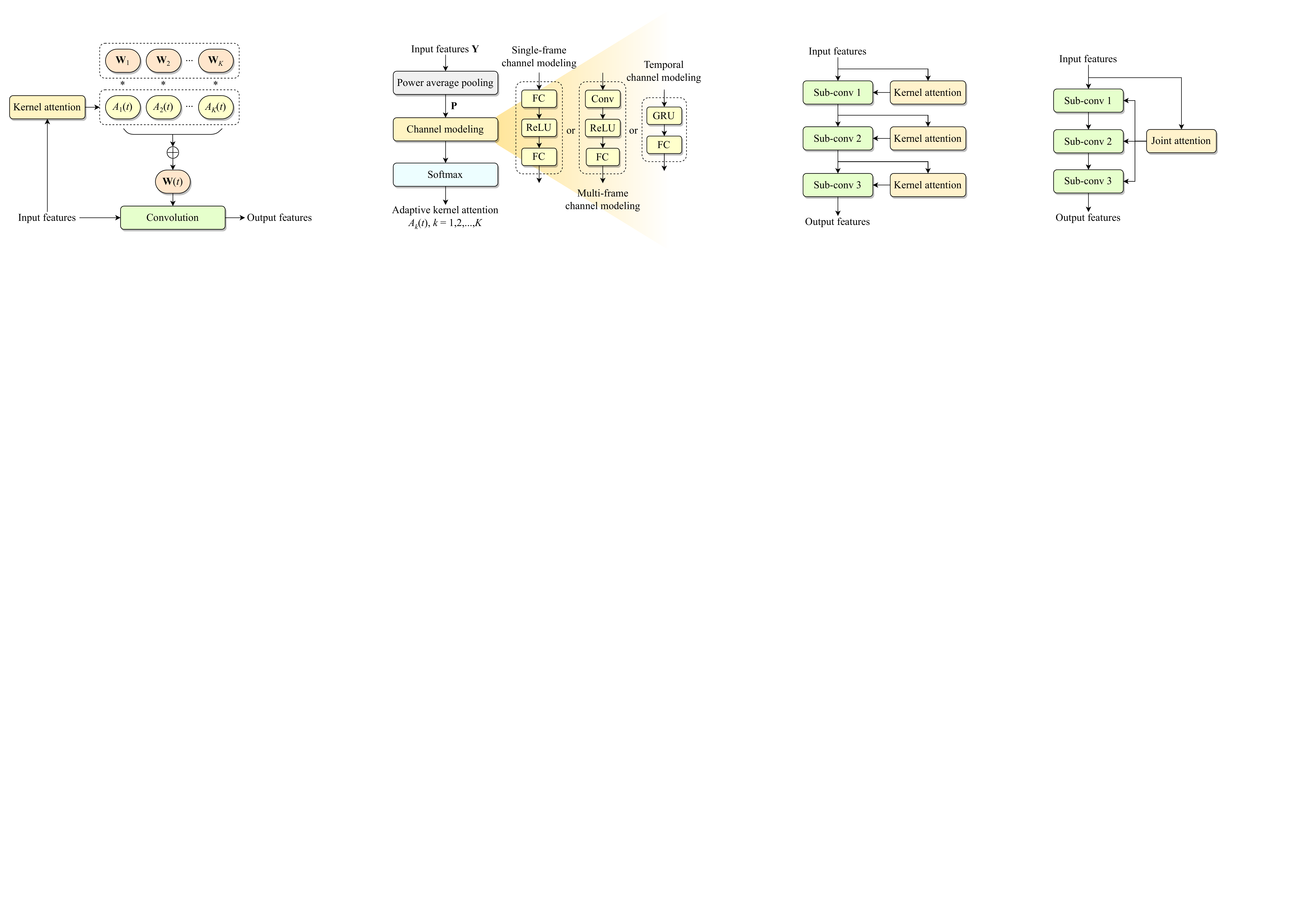}}
    \hfil
    \subfigure[]{\includegraphics[height=1.8in]{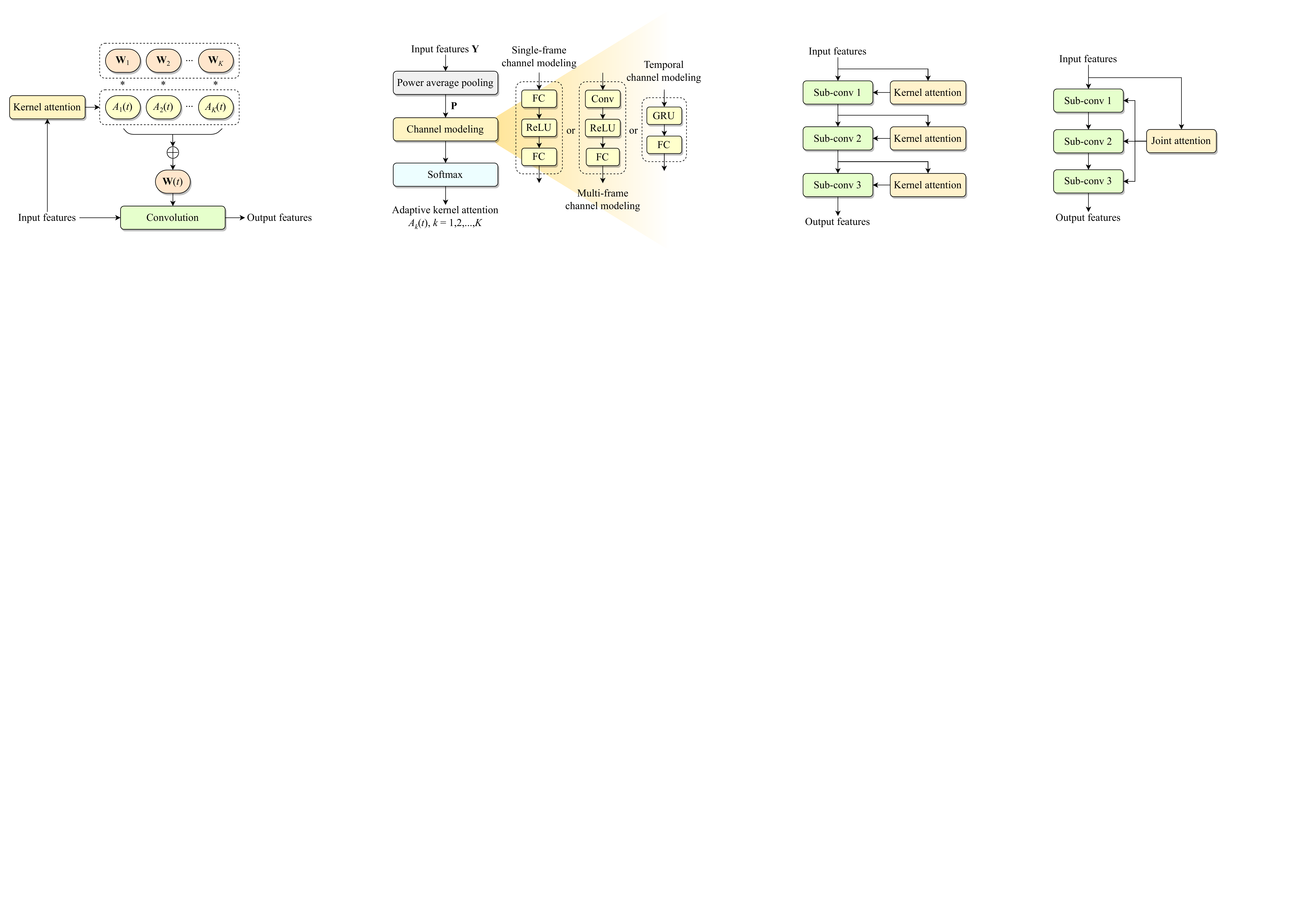}}
    \caption{Schematic illustration of adaptive convolution: (a) overall architecture of adaptive convolution layer, (b) architecture of kernel attention, (c) natural structure of a convolutional block with multiple sub-layers incorporating adaptive convolution, and (d) corresponding structure using joint multi-layer kernel attention.}
    \label{fig1}
\end{figure}

The attention mechanism is pivotal to the effectiveness of adaptive convolution, as it assigns attention weights to each parallel candidate kernel, determining the kernel composition for the current frame. As shown in Fig.~\ref{fig1}(b), kernel attention aggregates the input features $\mathbf{Y} \in \mathbb{R}^{C \times T \times F}$ along the frequency dimension through power average pooling to derive the descriptor $\mathbf{P} \in \mathbb{R}^{C \times T}$ for the time-channel energy distribution, where $C,T,F$ represent the number of channels, frames and frequency bins, respectively. The corresponding pooling is formulated as
\begin{equation}
    P\left(c,t\right) = \frac{1}{F} \sum_{f=0}^{F-1} Y^2 \left(c,t,f\right),
    \label{eq.4}
\end{equation}
where $c$ and $f$ denote the channel and frequency bin indices, respectively. $\mathbf{P}$ is subsequently processed through channel modeling and a softmax function to compute the time-varying attention weights $A_k(t)$ for each candidate kernel.

For causal convolution, an intuitive channel modeling approach is the single-frame channel modeling strategy. The compressed feature $\mathbf{P}$ is processed through an FC layer, a ReLU activation, and another FC layer sequentially, which essentially corresponds to the standard squeeze-and-excitation mechanism. Notably, the single-frame channel modeling approach may implicitly leverage historical information when adaptive convolution is used as a layer within the overall model. This is because the input features at each frame may already encode information from previous frames, depending on the structure of the preceding layers. Additionally, we explore a multi-frame channel modeling approach, which consists of a stack of a one-dimensional (1-D) convolution layer, a ReLU activation, and an FC layer. The 1-D convolution explicitly incorporates information from previous frames, thereby extending the receptive field of channel modeling. Inspired by TRA \cite{rong2024gtcrn}, a lightweight temporal channel attention mechanism, we adopt a similar strategy by using gated recurrent unit (GRU) to model channel features along the time dimension. Specifically, a GRU followed by an FC layer is employed for temporal channel modeling. This leads to better utilization of inter-frame correlations, thus yielding more efficient adaptive kernels, albeit at the cost of introducing more parameters compared to single-frame modeling.

Benefiting from the lightweight design of kernel attention, adaptive convolution introduces limited computational complexity to the model when the original convolution kernel scale is not particularly large, which is critical in practical applications. However, this comes with an increase in parameters, as each adaptive convolution layer contains $K$ parallel candidate convolution kernels, resulting in a convolutional parameter count that is $K$ times the original. Moreover, the additional computational load primarily arises from the kernel aggregation step. When the original model has large convolution kernels, the resulting computational burden can become considerable, imposing certain constraints on its applicability.

\subsection{Joint Attention \label{sec3.2}}
Efficient convolutional blocks, such as residual blocks \cite{he2016deep} and ConvNeXt blocks \cite{liu2022convnet}, typically consist of multiple depthwise and pointwise convolution layers. Replacing each sub-layer with adaptive convolution results in the natural structure depicted in Fig.~\ref{fig1}(c). Given the strong correlations and similarities among hidden features across sub-layers within an individual block, we propose a joint attention mechanism that generates kernel attention weights for multiple sub-layers simultaneously at the beginning of the block using a multi-head strategy. Fig.~\ref{fig1}(d) showcases an example of joint multi-layer attention applied to a block with three sub-layers. Specifically, in the channel modeling stage, the final FC layer is replaced with multiple parallel heads, each producing the kernel attention weights for a specific sub-layer. To enhance computational efficiency, we adopt a highly parallelized implementation by increasing the output channels of the FC layer to $N \times K$, where $N$ is the number of sub-layers. The output feature is then split along the channel dimension to derive the kernel attention weights for each sub-layer. The joint multi-layer attention simplifies the model structure by computing the adaptive kernels for all sub-convolutions within a single attention layer. Additionally, it slightly reduces the model size and MACs, particularly in configurations that use temporal channel modeling.

Inspired by ODConv \cite{li2022odconv}, the joint attention mechanism can be further extended to generate attention maps for other dimensions of the kernel space, including joint channel attention and joint spatial attention. Specifically, joint channel attention assigns time-varying attention weights $\mathbf{A}^\text{c}(t) \in \mathbb{R}^{C_{\text{in}}}$ and $\mathbf{A}^\text{f}(t) \in \mathbb{R}^{C_{\text{out}}}$ to the input channel and output channel dimensions of the kernel. Joint spatial attention generates time-varying weights $\mathbf{A}^\text{s} \in \mathbb{R}^{K_\text{t} \times K_\text{f}}$ for each spatial location of the convolution kernel with shape $(K_\text{t}, K_\text{f})$. For efficiency, all attention weights are shared across the candidate kernels. The kernel aggregation incorporating joint channel and spatial attention is formulated as
\begin{equation}
    \mathbf{W}\left(t\right) = \mathbf{A}^\text{s}\left(t\right) * \mathbf{A}^\text{c}\left(t\right) * \mathbf{A}^\text{f}\left(t\right) * \left[ \sum_{k=1}^{K} A_k\left(t\right) \mathbf{W}_k \right],
    \label{eq.5}
\end{equation}
where the attention weights are implicitly broadcasted to align with the shape of $\mathbf{W}_k$ before element-wise multiplication. These attention weights are generated by extending the kernel attention module with additional attention heads followed by sigmoid activation, similar to the design of joint multi-layer attention. These optional structures enable more precise adjustment of convolution kernels, enhancing their adaptability to the time-varying speech statistical characteristics.

\subsection{Multi-frame Parallelism \label{sec3.3}}
Adaptive convolution generates independent convolution kernels for each frame in a causal manner, preventing parallel computation across different frames of an utterance. While this limitation does not affect streaming inference, where frames are processed sequentially, it poses a challenge for parallel computing across all frames in an audio sample. In addition to explicitly implementing convolution through matrix multiplication, we propose two approaches to enable multi-frame parallelism. 

The first approach involves converting kernel aggregation into the aggregation of convolution outputs. During training, the convolution results for each candidate kernel are computed first, whose results are then aggregated using kernel attention to produce the final output through a weighted sum. This process is mathematically equivalent to adaptive kernel aggregation, as demonstrated by
\begin{equation}
    \begin{aligned}
    \mathbf{Z}\left(t\right) = \mathbf{W}\left(t\right) \circledast \mathbf{Y}\left(t\right) = \left[ \sum_{k=1}^{K} A_k\left(t\right) \mathbf{W}_k \right] \circledast \mathbf{Y}\left(t\right)  \\ = \sum_{k=1}^{K} A_k\left(t\right) \bigg[ \mathbf{W}_k \circledast \mathbf{Y}\left(t\right) \bigg] = \sum_{k=1}^{K} A_k\left(t\right) \mathbf{Z}_k\left(t\right),
    \end{aligned}
    \label{eq.6}
\end{equation}
where $\mathbf{Z}(t)$ and $\mathbf{Y}(t)$ denote the input and output features of the $t$-th frame, respectively, $\mathbf{Z}_k$ represents the output obtained by applying the $k$-th candidate kernel independently, which can be computed in parallel as usual, and $\circledast$ denotes the convolution operation. However, this method is restricted to the training stage due to the significant increase in computational complexity, approximately $K$ times the original.

The second approach leverages grouped convolution to simulate streaming inference while enabling multi-frame parallelism without increasing computational complexity. Inspired by DyConv \cite{chen2020dynamic}, which addresses parallelism issues for different images in a batch by merging the batch dimension into the channel dimension and assigning each image to a different group, we extend this concept to incorporate both batch and time dimensions. Specifically, an unfold operation is first performed along the time dimension of the input features with kernel size of $K_t$, where $K_t$ denotes the size of the convolution kernel along the time dimension. This operation extracts a sliding window of features for each time step of the convolution. Subsequently, the batch and time dimensions are merged into the channel dimension, forming an input tensor of shape $\left( 1, B \times T \times C, K_t, F \right)$, where $B$ is the batch size. The reshaped input is then divided into $B \times T$ groups, and convolution is performed with the assembled adaptive kernels. 

\section{AdaptCRN \label{sec4}}
Adaptive convolution achieves significant performance improvements with limited increases in computational complexity, making it an effective component for ultra-lightweight models designed for real-time applications, which typically impose strict constraints on computational resources. Additionally, in ultra-lightweight models, the inherently small convolution kernel size ensures that both the total parameter count and the added computational load remain within an acceptable range when adaptive convolution is incorporated. Therefore, we propose the adaptive convolution recurrent network (AdaptCRN), an ultra-lightweight model that leverages adaptive convolution to enhance performance while maintaining efficiency.

\subsection{Overall Architecture \label{sec4.1}}
\begin{figure}[!t]
    \centering
    \subfigure[]{\includegraphics[height=2.2in]{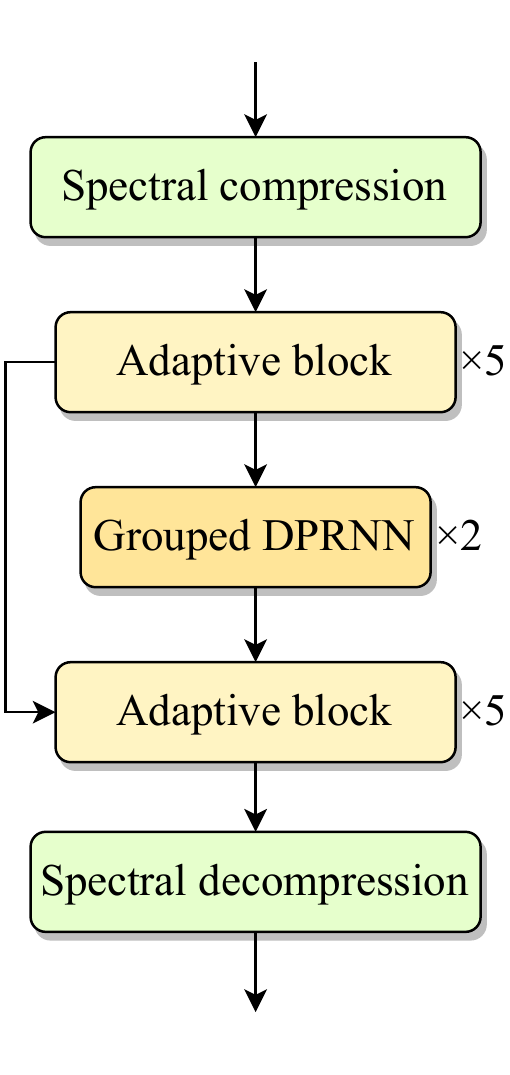}}
    \hfil
    \subfigure[]{\includegraphics[height=2.2in]{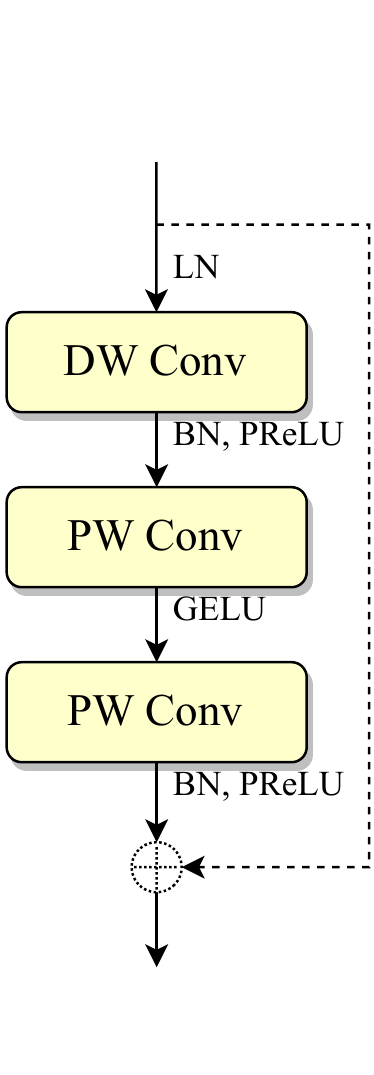}}
    \hfil
    \subfigure[]{\includegraphics[height=2.2in]{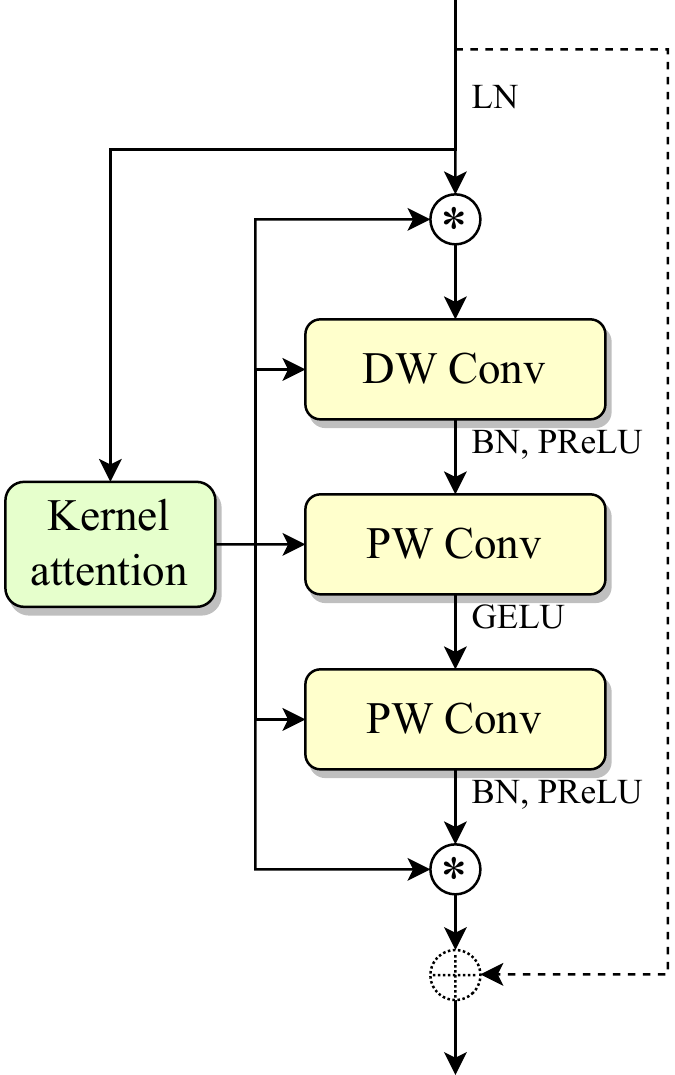}}
    \caption{The architecture of (a) AdaptCRN, (b) the basic block, and (c) the adaptive block.}
    \label{fig2}
\end{figure}

AdaptCRN works in the short-time Fourier transform (STFT) domain and takes the spectral magnitude as well as the real and imaginary parts as input. Its architecture consists of a spectral compression module, an encoder with 5 layers of adaptive blocks, 2 grouped DPRNN modules, a decoder with 5 layers of adaptive blocks, and a spectral decompression module, as shown in Fig.~\ref{fig2}(a). Skip connections are employed between corresponding layers of the encoder and decoder to facilitate optimization and alleviate information loss.

Grouped DPRNN \cite{rong2024gtcrn} combines grouped RNN \cite{gao2018efficient} and DPRNN \cite{le2021dpcrn,luo2020dual} to model spectral patterns in a single frame and inter-frame dependence within a certain frequency bin, while reducing the model complexity. Consistent with \cite{rong2024gtcrn}, we adopt grouped GRUs and eliminate representation rearrangement between time steps within the grouped RNN to improve efficiency. Furthermore, in our implementation, we remove the representation rearrangement after the grouped RNN, as the grouped RNN in DPRNN is directly followed by an FC layer, which inherently performs inter-group information fusion. Moreover, representation rearrangement is equivalent to swapping the rows of the FC layer's weight matrix, which does not alter the final output.

The model predicts the spectral magnitude mask rather than the complex ratio mask (CRM). Our observations indicate that for trained ultra-lightweight models, the imaginary part of CRM is usually close to 0, making it effectively equivalent to a magnitude mask. To enhance mask learning across different frequency bins, we employ a learnable sigmoid function \cite{yan2024lisennet,fu2021metricgan+} as the activation for the magnitude mask. Further details about the model are described in Secs.~\ref{sec4.2}, \ref{sec4.3}, \ref{sec4.4}.

\subsection{Spectral Compression \label{sec4.2}}
The spectral compression module comprises frequency dimension compression and dynamic range compression. For frequency dimension compression \cite{rong2024gtcrn}, low-frequency components below 2 kHz remain unaltered, while high-frequency bins above 2 kHz are downsampled using a triangular filter based on the equivalent rectangular bandwidth (ERB) scale. This retains crucial harmonic information in low-frequency bins while downsampling the frequency dimension. Dynamic range compression equalizes spectral energy, clarifies the spectral structure, and facilitates the model's feature extraction. Specifically, logarithmic operation is applied to compress spectral magnitudes, while a magnitude power law is used to compress the real and imaginary parts. The specific spectral compression is formulated as
\begin{equation}
    X_{\text{mag}}^{\text{c}} = \log_{10} \left( f_\text{ERB} \left( \left| X \right| \right) \right),
    \label{eq.7}
\end{equation}
\begin{equation}
    X_{\text{r/i}}^{\text{c}} =  f_\text{ERB} \left( \frac{X_\text{r/i}}{\left|X\right|^{0.7}} \right),
    \label{eq.8}
\end{equation}
where $X, X_\text{r}, X_\text{i}$ denote the noisy spectrum and its real and imaginary parts, respectively, the superscript $\text{c}$ represents the compressed features, and $f_\text{ERB}(\cdot)$ denotes the frequency dimension downsampling operation based on the ERB scale. For the compressed features, we employ the SFE module from \cite{rong2024gtcrn} to integrate information from adjacent subbands into the channel dimension. The spectral decompression module upsamples the frequency dimension of the single-channel output of the decoder to produce the predicted mask. The decompression is achieved by applying the transpose of the frequency dimension downsampling matrix, rather than using a learnable transformation.

\subsection{Adaptive Block \label{sec4.3}}
Inverted residual blocks, ConvNeXt blocks, and StarNet blocks are modern CNN designs that have been proven to be lightweight and efficient. These blocks share a similar structure comprising one depthwise convolution (DW Conv) layer and two pointwise convolution (PW Conv) layers. Inspired by these designs, we propose the basic block for AdaptCRN, as depicted in Fig.~\ref{fig2}(b). The input features are first processed by layer normalization (LN) along the channel and frequency dimensions, which smooths the energy distribution and preserves spectral patterns across different channels and subbands. The element-wise affine transformation in LN acts as a set of fixed filters to preprocess spectral features. Subsequently, the features pass through a DW Conv layer followed by two PW Conv layers in sequence. Both the DW Conv and the second PW Conv are followed by batch normalization (BN) and parametric ReLU (PReLU) activation, while Gaussian error linear unit (GELU) activation is applied between the two PW Conv layers. We also consider the star operation from StarNet as a replacement for GELU, which is defined as $f_\text{ReLU6}\left( x \right) * x$, with $x$ the input features and $f_\text{ReLU6}(\cdot)$ the ReLU6 activation function. However, experimental results do not indicate a performance advantage over GELU. Additionally, when frequency dimension downsampling is not required and the number of input and output channels is equal, skip connections are employed to directly add the input features to the output features, mitigating information loss in deeper layers.

By incorporating adaptive convolution into the basic block, we derive the adaptive block, as illustrated in Fig~\ref{fig2}(c). This block employs joint attention to generate both the input/output temporal channel attention maps and the adaptive kernel attention weights for the three sub-layers. For efficiency, the joint channel attention maps are applied to the input and output features of the three sub-layers, rather than to each individual kernel within those layers as described in Section~\ref{sec3.2}.

Adaptive convolution introduces high nonlinearity to the model, which is particularly evident when the activation function between the two PW Conv layers is removed. Typically, two PW Conv layers without activation can be reparameterized into a single layer, expressed as
\begin{equation}
    \mathbf{W}^{(2)}\left(\mathbf{W}^{(1)}\mathbf{x}\right) = \left(\mathbf{W}^{(2)}\mathbf{W}^{(1)}\right)\mathbf{x},
    \label{eq.9}
\end{equation}
where $\mathbf{W}^{(1)},\mathbf{W}^{(2)}$ denote the kernels of the two PW Conv layers, respectively, and the bias term is omitted for simplicity, as its inclusion does not alter the result. However, with adaptive convolution, the reparameterization changes to 
\begin{equation}
    \begin{aligned}
    \left(\sum_{k=1}^{K} A_k^{(2)} \mathbf{W}_k^{(2)}\right) \left[ \left(\sum_{k=1}^{K} A_k^{(1)} \mathbf{W}_k^{(1)}\right) \mathbf{x} \right] \\= \left[\sum_{j=1}^{K}\sum_{i=1}^{K} A_j^{(2)}A_i^{(1)} \left(\mathbf{W}_j^{(2)}\mathbf{W}_i^{(1)}\right)\right]\mathbf{x}, 
    \end{aligned}
    \label{eq.10}
\end{equation}
making two adaptive PW Conv layers with $K$ candidate kernels equivalent to a single layer with $K^2$ kernels. Since the resulting $K^2$ kernels and their corresponding kernel attention weights are derived from two groups of $K$ kernels, they reside in a low-rank space relative to directly optimizing $K^2$ independent kernels. Consequently, the reparameterized results may exhibit slightly reduced performance compared to directly optimizing a single layer with $K^2$ kernels.

\subsection{Loss Function \label{sec4.4}}
We apply the negative scale invariant SNR (SI-SNR) \cite{le2019sdr} loss and the power-compressed specturm loss as the loss functions, defined as
\begin{equation}
    \mathcal{L}_{\text{SI-SNR}} \left(\hat{s}, s\right) = -\log_{10}\left(\frac{\left\|s_{t}\right\|^{2}}{\left\|\hat{s}-s_{t}\right\|^{2}}\right); s_{t} = \frac{\langle\hat{s}, s\rangle s}{\left\|s\right\|^{2}}, 
    \label{eq.11}
\end{equation}
\begin{equation}
    \mathcal{L}_{\text{mag}}\left(\hat{S}, S\right) = \operatorname{MSE}\left(|\hat{S}|^{0.3}, |S|^{0.3}\right),
    \label{eq.12}
\end{equation}
\begin{equation}
    \mathcal{L}_{\text{real/imag}}\left(\hat{S}, S\right) = \operatorname{MSE}\left(\frac{\hat{S}_{\text{r/i}}}{|\hat{S}|^{0.7}}, \frac{S_{\text{r/i}}}{|S|^{0.7}}\right),
    \label{eq.13}
\end{equation}
where $s$ and $\hat{s}$ represent clean and enhanced waveforms, $S$ and $\hat{S}$ are their corresponding spectrograms, the subscripts $\text{r, i}$ represent the real and imaginary parts of the spectrograms, respectively, $\langle \cdot,\cdot \rangle$ denotes the inner product operator, and $\text{MSE}(\cdot,\cdot)$ represents the mean squared error (MSE). The overall loss function for model training is given by
\begin{equation}
    \mathcal{L} = \lambda_1 \mathcal{L}_{\text{SI-SNR}} + \lambda_2\mathcal{L}_{\text{mag}} + \lambda_3\left(\mathcal{L}_{\text{real}} + \mathcal{L}_{\text{imag}}\right), 
    \label{eq.14}
\end{equation}
where $\lambda_1, \lambda_2, \lambda_3$ are the empirical weights.

\section{Experimental Setup \label{sec5}}
\subsection{Datasets \label{sec5.1}}
We evaluate our proposed methods on two datasets. The first dataset is constructed using clean and noise samples from the 5th Deep Noise Suppression Challenge (DNS5) dataset \cite{dubey2024dns5}. Clean and noise clips are mixed at random SNR levels ranging from -5 dB to 15 dB and scaled to random amplitudes ranging from 0.01 to 0.99 of the full scale to improve the model's adaptation to magnitude dynamics. Approximately 200 hours of 10-second noisy-clean paired samples are generated for training, while 1,000 samples are generated for validation and testing, respectively. All utterances are resampled to 16 kHz for the experiments.

Additionally, we use the Voicebank+DEMAND dataset \cite{valentini2016investigating} to further evaluate the proposed ultra-lightweight model. It is a widely used benchmark for SE tasks, which contains paired clean and pre-mixed noisy speech. The training set comprises 11,572 utterances from 28 speakers, while the test set consists of 872 utterances from 2 speakers. SNR levels for training are set to 0, 5, 10, and 15 dB, and for testing, they are set to 2.5, 7.5, 12.5, and 17.5 dB. All utterances are downsampled from 48 kHz to 16 kHz.

\subsection{Baseline Models and Ablation Study \label{sec5.2}}
To evaluate the effectiveness of the proposed adaptive convolution, we integrate it into diverse models with varying scales and architectures, including DPCRN \cite{le2021dpcrn} of different sizes, DCCRN \cite{hu2020dccrn}, GTCRN \cite{rong2024gtcrn}, and LiSenNet \cite{yan2024lisennet}. DPCRN, as a well-established baseline, serves as a foundational structure for modern CRN-based SE tasks. We utilize three scaled versions of DPCRN, denoted as DPCRN-light, DPCRN-middle, and DPCRN-large.

Ablation experiments are conducted on DPCRN-light to verify the effectiveness of the proposed channel modeling techniques and the joint channel and spatial attention mechanism. Additionally, we validate the advantage of frame-level kernel adjustment by comparing with utterance-level global dynamic convolution, which employs non-causal attention with global pooling. Furthermore, we investigate the impact of kernel generation configurations on performance, focusing on the effectiveness of the temperature annealing training strategy and comparing candidate kernel aggregation via softmax-based attention with directly generating kernel parameters through the attention module.

For the proposed AdaptCRN, ablation experiments are performed to validate the contributions of incorporating adaptive convolution, joint channel attention, joint multi-layer kernel attention, and dynamic range compression of input features. The nonlinearity introduced by adaptive convolution is validated through experiments on cascaded PW Conv layers and their equivalent single-layer structure. Moreover, by comparing AdaptCRN with DPCRN-light \cite{le2021dpcrn}, DeepFilterNet \cite{schroter2022deepfilternet}, GTCRN \cite{rong2024gtcrn}, LiSenNet \cite{yan2024lisennet}, ULCNet \cite{shetu2024ultra}, and FSPEN \cite{yang2024fspen}, we demonstrate the superior performance of the proposed model.

\subsection{Implementation Details \label{sec5.3}}
The STFT is computed with a segment length of 32 ms, an overlap of 50\%, a fast Fourier transform (FFT) length of 512, and a square-root Hanning window for analysis and synthesis. For adaptive convolution, the number of candidate kernels $K$ is set to 8, unless otherwise specified. The number of hidden channels for channel modeling is set to 32, while the convolution kernel size of the 1D-Conv used in multi-frame channel modeling is set to 3. Optional joint channel and spatial attention mechanisms are not used unless specified in ablation experiments. During training, we do not apply the temperature annealing strategy for the softmax function described in Section \ref{sec2.1}, except in the specific ablation experiment. This is because our experiments show that it does not enhance kernel optimization in adaptive convolution, as detailed in Section \ref{sec6.1.2}.

For the three scaled versions of DPCRN, the number of DPRNN blocks is set to 2. The convolution kernel sizes $(K_t,K_f)$ for the encoder layers are configured as \{(2,5), (2,3), (2,3), (2,3), (2,3)\}, with frequency dimension strides of 2 for the first three layers and 1 for the last two layers. The number of output channels per layer is set to \{16, 16, 16, 32, 32\} for DPCRN-light, \{32, 32, 32, 64, 64\} for DPCRN-middle and \{32, 32, 32, 64, 128\} for DPCRN-large. The hidden dimensions of DPRNN are set to 32, 64, and 128, respectively. The decoder mirrors the configuration of the encoder. For GTCRN, LiSenNet, and DCCRN, the model hyperparameter settings remain consistent with their respective original papers, with the exception that the lookahead in the decoder of DCCRN is removed to ensure causality. For GTCRN, its main branch of the GT-Conv block contains three sub-layers and a TRA layer for generating the temporal channel attention map. When the vallina convolution are replaced with adaptive convolution, we employ joint multi-layer kernel attention to generate kernel attention weights for the three sub-layers, and the TRA is replaced with joint output channel attention to produce the temporal channel attention map.

For AdaptCRN, as detailed in Section~\ref{sec4.2}, the amplitude, real part, and imaginary part of the noisy spectrum are fed into the spectral compression module. The first 65 low-frequency bands (below 2 kHz) remain unaltered, while the 192 high-frequency bands (above 2 kHz) are mapped to 64 bands using the ERB-based filter described in Section~\ref{sec4.2}. 9-channel, 129-dimensional features per frame are obtained using SFE with a kernel size of 3. The hidden channels of all adaptive blocks are set to 16, except in the final block, where the hidden channels between the two PW Convs are reduced to 4. In the encoder, the convolution kernel sizes of the first two layers are (1,5) with a frequency dimension stride of 2, while the last three layers use kernel sizes of (3,3) with a stride of 1. In the decoder, the last two layers employ depthwise transposed convolutions, with the remaining settings matching the encoder. The grouped DPRNN consists of 2 groups, with a frequency dimension of 33. The hidden channels of the intra-frame and inter-frame RNNs are set to 8 and 16, respectively. The single-channel output of the final adaptive block undergoes spectral decompression followed by a learnable sigmoid function to generate the amplitude mask.

All models are trained using the Adam optimizer with an initial learning rate of 0.001. For DNS5 dataset, 10,000 samples are randomly selected for training in each epoch, with a batch size of 8. For DPCRN-light, GTCRN, LiSenNet, DCCRN, and AdaptCRN, the learning rate is halved at epochs 120, 150, 170, 180, 190, and 200. For DPCRN-middle and DPCRN-large, the learning rate is halved if the validation loss does not decrease for 10 consecutive epochs. For the Voicebank+DEMAND dataset, a cosine annealing scheduler is employed for learning rate decay over 300 epochs with a minimum learning rate of 1e-5. All models are trained with the cost function described in Section~\ref{sec4.4}, with $\lambda_1=0.01$, $\lambda_2=0.7$, and $\lambda_3=0.3$.

\subsection{Evaluation Metrics \label{sec5.4}}
We employ a set of commonly used speech objective metrics for evaluation, including SI-SNR \cite{le2019sdr}, wide-band perceptual evaluation of speech quality (PESQ) \cite{rix2001perceptual}, short-time objective intelligibility (STOI) \cite{taal2010short} and its extended version (ESTOI) \cite{jensen2016algorithm}, and deep noise suppression mean opinion score (DNSMOS) \cite{reddy2022dnsmos}. DNSMOS is a non-intrusive speech quality metric, which provides three scores including speech quality (SIG), background noise quality (BAK), and overall quality (OVRL). Higher values indicate better performance for all metrics.

\section{Experimental Results \label{sec6}}
\subsection{Experimental Results for adaptive convolution \label{sec6.1}}

\subsubsection{Ablation study results for kernel attention \label{sec6.1.1}}
Table~\ref{Table1} presents the results of ablation experiments conducted on DPCRN-light, exploring the configuration of kernel attention for adaptive convolution. In these experiments, the number of candidate convolution kernels is fixed at 8. The table reports trainable parameter numbers, MACs, and evaluation results on DNS5 test set for three channel modeling approaches: single-frame, multi-frame, and temporal channel modeling. The results demonstrate that the adaptive convolution with any channel modeling approach significantly outperforms the vanilla convolution across all metrics, achieving approximately a 0.1 gain in PESQ and a 0.05 improvement in DNSMOS-OVRL, with only a modest increase in computational complexity. Among the three approaches, temporal channel modeling introduces more parameters but achieves the most substantial performance gains. This indicates that utilizing a GRU for temporal channel modeling effectively leverages historical information, enabling the generation of more suitable adaptive kernels for each frame.

Building on temporal channel modeling, we further investigate the impact of incorporating optional joint channel and spatial attention. The results in Table~\ref{Table1} reveal that adding joint channel attention alone yields an improvement in performance, with minimal additional parameters and computational cost. In contrast, adding spatial attention does not provide noticeable benefits. This suggests that joint channel attention can further calibrate signal energy at the frame level, enhancing the model's performance.

Moreover, we evaluate the performance of utterance-level global dynamic convolution, where kernel attention is derived using global average power pooling across the entire spectrogram followed by a squeeze-and-excitation mechanism. The results show that its performance is comparable to adaptive convolution with single-frame channel modeling but is inferior to the other two channel modeling approaches, even though it employs a highly non-causal mechanism. This highlights the importance of frame-level kernel adjustments, which are more effective for modeling the time-varying feature of speech signals.

\begin{table*}[t]
\renewcommand{\arraystretch}{1.1}
\caption{\label{Table1} Ablation study results for kernel attention on DNS5 test set.}
\centering
\begin{tabular}{p{4.3cm}<{\centering}p{1.2cm}<{\centering}p{1.2cm}<{\centering}p{1.2cm}<{\centering}p{1.2cm}<{\centering}p{1.2cm}<{\centering}p{1.2cm}<{\centering}p{1.2cm}<{\centering}p{1.2cm}<{\centering}}
\Xhline{1pt}
\multirow{2}*{} & \multirow{2}*{Para. (K)} & \multirow{2}*{MACs (M)} & \multirow{2}*{SI-SNR} & \multirow{2}*{ESTOI} & \multirow{2}*{PESQ} & \multicolumn{3}{c}{DNSMOS} \\ \cline{7-9} ~ & ~ & ~ & ~ & ~ & ~ & OVRL & SIG & BAK \\
\hline
Noisy & - & - & 5.366 & 0.625 & 1.391 & 2.147 & 2.918 & 2.340 \\
Vanilla convolution & 80.78  & 194.57 & 14.431 & 0.752 & 2.313 & 2.910 & 3.197 & 4.018 \\
Globle dynamic convolution & 358.62 & 194.88 & 15.060 & 0.758 & 2.410 & 2.924 & 3.210 & 4.023 \\
\hline
Single-frame channel modeling & 358.62 & 210.45 & 14.924 & 0.761 & 2.390 & 2.946 & 3.229 & 4.034 \\
Multi-frame channel modeling & 378.21 & 211.68 & 14.956 & 0.762 & 2.406 & 2.957 & 3.239 & 4.038 \\
Temporal channel modeling & 410.21 & 213.80 & 15.147 & 0.765 & 2.427 & 2.964 & 3.246 & 4.040 \\
\hline
+ Joint channel attention & 426.71 & 214.84 & 15.177 & 0.766 & 2.445 & 2.974 & 3.257 & 4.043 \\
+ Joint spatial attention & 412.45 & 213.94 & 15.133 & 0.766 & 2.427 & 2.971 & 3.252 & 4.045 \\
+ Joint channel/spatial attention & 428.95 & 214.98 & 15.192 & 0.766 & 2.445 & 2.970 & 3.253 & 4.040 \\
\Xhline{1pt}
\end{tabular}
\end{table*}

\begin{table*}[t]
\renewcommand{\arraystretch}{1.1}
\caption{\label{Table2} Ablation study results for the detailed configuration of kernel aggregation on DNS5 test set.}
\centering
\begin{tabular}{p{4.7cm}<{\centering}p{1.2cm}<{\centering}p{1.2cm}<{\centering}p{1.2cm}<{\centering}p{1.2cm}<{\centering}p{1.2cm}<{\centering}p{1.2cm}<{\centering}p{1.2cm}<{\centering}p{1.2cm}<{\centering}}
\Xhline{1pt}
& \multirow{2}*{Para. (K)} & \multirow{2}*{MACs (M)} & \multirow{2}*{SI-SNR} & \multirow{2}*{ESTOI} & \multirow{2}*{PESQ} & \multicolumn{3}{c}{DNSMOS} \\ \cline{7-9} ~ & ~ & ~ & ~ & ~ & ~ & OVRL & SIG & BAK \\
\hline
Noisy & - & - & 5.366 & 0.625 & 1.391 & 2.147 & 2.918 & 2.340 \\
\hline
Adaptive convolution ($K = 8$) & 410.21 & 213.80 & 15.147 & 0.765 & 2.427 & 2.964 & 3.246 & 4.040 \\
Adaptive convolution ($K = 32$) & 1325.90 & 271.51 & 15.278 & 0.769 & 2.453 & 2.982 & 3.266 & 4.043 \\
\hline
w/ temperature annealing & 410.21 & 213.80 & 15.091 & 0.765 & 2.425 & 2.963 & 3.243 & 4.043 \\
\hline
w/o softmax & 410.22 & 213.81 & 15.086 & 0.764 & 2.429 & 2.964 & 3.244 & 4.043 \\
w/o softmax \& dimensionality reduction & 1353.17 & 277.25 & 15.139 & 0.767 & 2.451 & 2.969 & 3.252 & 4.042 \\
\Xhline{1pt}
\end{tabular}
\end{table*}

\subsubsection{Detailed configuration of kernel generation \label{sec6.1.2}}
Table~\ref{Table2} first reports the performance of DPCRN-light with adaptive convolution using 32 candidate kernels. While this achieves a slight improvement over our default configuration of $K=8$, expanding the kernel space comes at the cost of significantly more parameters and higher computational complexity. Based on our experiments, 8 candidate kernels provide a reasonable configuration, offering a good trade-off between performance and efficiency. We further examine the temperature annealing strategy for training kernel attention, as recommended in DyConv for CV tasks (see Section \ref{sec2.1}). As shown in Table~\ref{Table2}, applying this strategy to adaptive convolution yields essentially the same results as without it, suggesting that temperature annealing may not be effective for SE models.

Additionally, we evaluate the kernel generation approach recommended in DGCN, where the output of the final FC layer of the attention module is directly used as the convolution kernel parameters. Conceptually, this is closely related to our approach: for each frame, our attention mechanism produces a $K$-dimensional weight vector, which aggregates the $K$ candidate kernels (equivalent to a $K \times D$ matrix, with $D$ denoting the flattened kernel size) into the final kernel. This process is equivalent to using a $K \times D$ FC layer to generate the kernel parameters. However, there are two important differences. First, the softmax normalization forms probability-like weights, each ranging from 0 to 1 and collectively summing to one. DyConv suggests this constraint facilitates kernel learning. Second, the use of kernel attention weights is equivalent to reducing the hidden feature dimension (set to 32) to $K$ (set to 8) before mapping it to the kernel size, thereby introducing a dimensionality reduction step within the attention module.

Table~\ref{Table2} also presents the results of removing the softmax constraint (replacing it with PReLU). Performance remains essentially unchanged, indicating that DyConv's insight regarding the benefit of attention score normalization does not hold in SE tasks. We further test directly using the output of the final FC layer as convolution kernel parameters, i.e., removing both the softmax constraint and the dimensionality reduction to $K$. This greatly increases parameter count and computational cost, but the performance improvement is very limited. This demonstrates that employing a low hidden dimension in the FC layer is an effective architectural choice.

Overall, these results provide deeper insight into the kernel generation, showing that certain conclusions drawn in CV do not transfer well to SE. Notably, we retain the kernel aggregation formulation and the softmax constraint on attention weights, as they provide improved interpretability for the kernel generation of adaptive convolution and support our subsequent analysis of the correlation between kernel allocation and speech features.

\begin{table*}[t]
\renewcommand{\arraystretch}{1.1}
\caption{\label{Table3} Generalization performance of adaptive convolution on DNS5 test set.}
\centering
\begin{tabular}{p{2cm}<{\centering}p{1.9cm}<{\centering}p{1.2cm}<{\centering}p{1.2cm}<{\centering}p{1.2cm}<{\centering}p{1.2cm}<{\centering}p{1.2cm}<{\centering}p{1.2cm}<{\centering}p{1.2cm}<{\centering}p{1.2cm}<{\centering}}
\Xhline{1pt}
\multirow{2}*{Model} & \multirow{2}*{Convolution} & \multirow{2}*{Para. (K)} & \multirow{2}*{MACs (M)} & \multirow{2}*{SI-SNR} & \multirow{2}*{ESTOI} & \multirow{2}*{PESQ} & \multicolumn{3}{c}{DNSMOS} \\ \cline{8-10} & ~ ~ & ~ & ~ & ~ & ~ & ~ & OVRL & SIG & BAK \\
\hline
Noisy & - & - & - & 5.366 & 0.625 & 1.391 & 2.147 & 2.918 & 2.340 \\
\hline
\multirow{2}*{GTCRN} & Vanilla & 23.67 & 33.83 & 13.552 & 0.727 & 2.130 & 2.838 & 3.124 & 3.992 \\
~ & Adaptive & 117.36 & 40.75 & 14.296 & 0.747 & 2.292 & 2.887 & 3.172 & 4.013 \\
\hline
\multirow{2}*{LiSenNet} & Vanilla & 36.78 & 55.77 & 14.352 & 0.742 & 2.244 & 2.897 & 3.180 & 4.018 \\
~ & Adaptive & 198.94 & 62.57 & 15.003 & 0.759 & 2.369 & 2.963
 & 3.240 & 4.045 \\
\hline
\multirow{2}*{DPCRN-light} & Vanilla & 80.78  & 194.57 & 14.431 & 0.752 & 2.313 & 2.910 & 3.197 & 4.018 \\
~ & Adaptive & 410.21 & 213.80 & 15.147 & 0.765 & 2.427 & 2.964 & 3.246 & 4.040 \\
\hline
\multirow{2}*{DPCRN-middle} & Vanilla & 300.81 & 734.31 & 15.569 & 0.775 & 2.521 & 2.997 & 3.281 & 4.047 \\
~ & Adaptive & 1440.29 & 809.77 & 15.948 & 0.782 & 2.582 & 3.030 & 3.311 & 4.059 \\
\hline
\multirow{2}*{DPCRN-large} & Vanilla & 787.15 & 1716.93 & 15.952 & 0.784 & 2.605 & 3.020 & 3.304 & 4.051 \\
~ & Adaptive & 2455.01 & 1829.54 & 16.139 & 0.787 & 2.621 & 3.040 & 3.324 & 4.054 \\
\hline
\multirow{2}*{DCCRN} & Vanilla & 3671.05 & 5601.27 & 14.870 & 0.768 & 2.458 & 2.989 & 3.273 & 4.043 \\
~ & Adaptive & 22449.07 & 8235.08 & 15.208 & 0.777 & 2.526 & 3.027 & 3.309 & 4.053 \\
\Xhline{1pt}
\end{tabular}
\end{table*}

\begin{table*}[t]
\renewcommand{\arraystretch}{1.1}
\caption{\label{Table4} Ablation study results for AdaptCRN on DNS5 test set.}
\centering
\begin{tabular}{p{4.3cm}<{\centering}p{1.2cm}<{\centering}p{1.2cm}<{\centering}p{1.2cm}<{\centering}p{1.2cm}<{\centering}p{1.2cm}<{\centering}p{1.2cm}<{\centering}p{1.2cm}<{\centering}p{1.2cm}<{\centering}}
\Xhline{1pt}
& \multirow{2}*{Para. (K)} & \multirow{2}*{MACs (M)} & \multirow{2}*{SI-SNR} & \multirow{2}*{ESTOI} & \multirow{2}*{PESQ} & \multicolumn{3}{c}{DNSMOS} \\ \cline{7-9} ~ & ~ & ~ & ~ & ~ & ~ & OVRL & SIG & BAK \\
\hline
Noisy & - & - & 5.366 & 0.625 & 1.391 & 2.147 & 2.918 & 2.340 \\
AdaptCRN & 134.51 & 40.80 & 14.892 & 0.759 & 2.387 & 2.939 & 3.219 & 4.037 \\
\hline
w/o Adaptive convolution & 29.44 & 33.67 & 13.826 & 0.736 & 2.192 & 2.872 & 3.158 & 4.008 \\
w/o Joint channel attention & 124.68 & 40.18 & 14.718 & 0.756 & 2.352 & 2.925 & 3.207 & 4.031 \\
w/o Joint multi-layer kernel attention & 169.21 & 43.19 & 14.803 & 0.759 & 2.386 & 2.939 & 3.221 & 4.036 \\
w/o Dynamic range compression & 134.51 & 40.80 & 14.695 & 0.755 & 2.342 & 2.929 & 3.211 & 4.031 \\
\hline
DW(8)-PW(8)-skip-PW(8) & 134.51 & 40.01 & 14.672 & 0.755 & 2.358 & 2.931 & 3.212 & 4.034 \\
DW(8)-PW(8) & 112.88 & 32.06 & 14.551 & 0.754 & 2.343 & 2.926 & 3.207 & 4.031 \\
DW(8)-PW(64) & 255.00 & 41.41 & 14.826 & 0.759 & 2.390 & 2.942 & 3.221 & 4.039 \\
\Xhline{1pt}
\end{tabular}
\end{table*}

\subsubsection{Generalization capability of adaptive convolution \label{sec6.1.3}}
We replace the vanilla convolutions with the adaptive convolutions in various CNN-based models of different structures and scales to evaluate the generalization capability of adaptive convolution. The results, as shown in Table~\ref{Table3}, demonstrate that adaptive convolution significantly improves the performance across all metrics, particularly in lightweight models. Beyond the improvements observed in DPCRN-light, adaptive convolution introduces approximately a 0.16 PESQ gain and a 0.05 DNSMOS-OVRL gain in GTCRN, as well as a PESQ gain exceeding 0.12 and a DNSMOS-OVRL gain over 0.06 in LiSenNet. These substantial improvements can be attributed to adaptive convolution's ability to expand the originally limited parameter space of lightweight models, thereby significantly enhancing their representation capacity.

However, as the model size increases, the improvements introduced by adaptive convolution become less pronounced. For DPCRN-middle and DCCRN, it achieves approximately a 0.06 PESQ gain and a 0.03 DNSMOS-OVRL gain. These improvements are relatively smaller compared to those observed in lightweight models. For DPCRN-large, the improvements are minimal, with only about a 0.2 dB SI-SNR gain, a 0.02 PESQ gain, and a 0.02 DNSMOS-OVRL gain. The diminishing improvement can be attributed to the extensive parameter space inherent in larger models. Specifically, the convolutional kernels in such models already include filters capable of adapting to diverse spectral features, reducing the benefit of additional candidate kernels.

Moreover, as the network size increases, the additional storage and computational overhead introduced by adaptive convolution becomes more substantial. This is primarily due to the increase in the overall size of the convolution kernels in larger models. The impact is particularly pronounced in the large DCCRN model, where convolution operations account for over 97\% of the total MACs. In this case, the additional cost introduced by adaptive convolution becomes prohibitively high, rendering the method less suitable for these architectures. These findings indicate that the applicability and effectiveness of the proposed adaptive convolution are limited in large-scale models.

\subsection{Experimental Results for AdaptCRN \label{sec6.2}}

\begin{table*}[t]
\renewcommand{\arraystretch}{1.1}
\caption{\label{Table5} Performance of AdaptCRN on DNS5 test set.}
\centering
\begin{tabular}{p{4.3cm}<{\centering}p{1.2cm}<{\centering}p{1.2cm}<{\centering}p{1.2cm}<{\centering}p{1.2cm}<{\centering}p{1.2cm}<{\centering}p{1.2cm}<{\centering}p{1.2cm}<{\centering}p{1.2cm}<{\centering}}
\Xhline{1pt}
\multirow{2}*{Model} & \multirow{2}*{Para. (K)} & \multirow{2}*{MACs (M)} & \multirow{2}*{SI-SNR} & \multirow{2}*{ESTOI} & \multirow{2}*{PESQ} & \multicolumn{3}{c}{DNSMOS} \\ \cline{7-9} ~ & ~ & ~ & ~ & ~ & ~ & OVRL & SIG & BAK \\
\hline
Noisy & - & - & 5.366 & 0.625 & 1.391 & 2.147 & 2.918 & 2.340 \\
DPCRN-light \cite{le2021dpcrn} & 80.78  & 194.57 & 14.431 & 0.752 & 2.313 & 2.910 & 3.197 & 4.018 \\
GTCRN \cite{rong2024gtcrn} & 23.67 & 33.83 & 13.552 & 0.727 & 2.130 & 2.838 & 3.124 & 3.992 \\
GTCRN-Adaptive & 117.36 & 40.75 & 14.296 & 0.747 & 2.292 & 2.887 & 3.172 & 4.013 \\
LiSenNet \cite{yan2024lisennet} & 36.78 & 55.77 & 14.352 & 0.742 & 2.244 & 2.897 & 3.180 & 4.018 \\
AdaptCRN & 134.51 & 40.80 & 14.892 & 0.759 & 2.387 & 2.939 & 3.219 & 4.037 \\
\Xhline{1pt}
\end{tabular}
\end{table*}

\begin{figure*}[!t]
    \centering
    \subfigure[]{\includegraphics[height=1.3in]{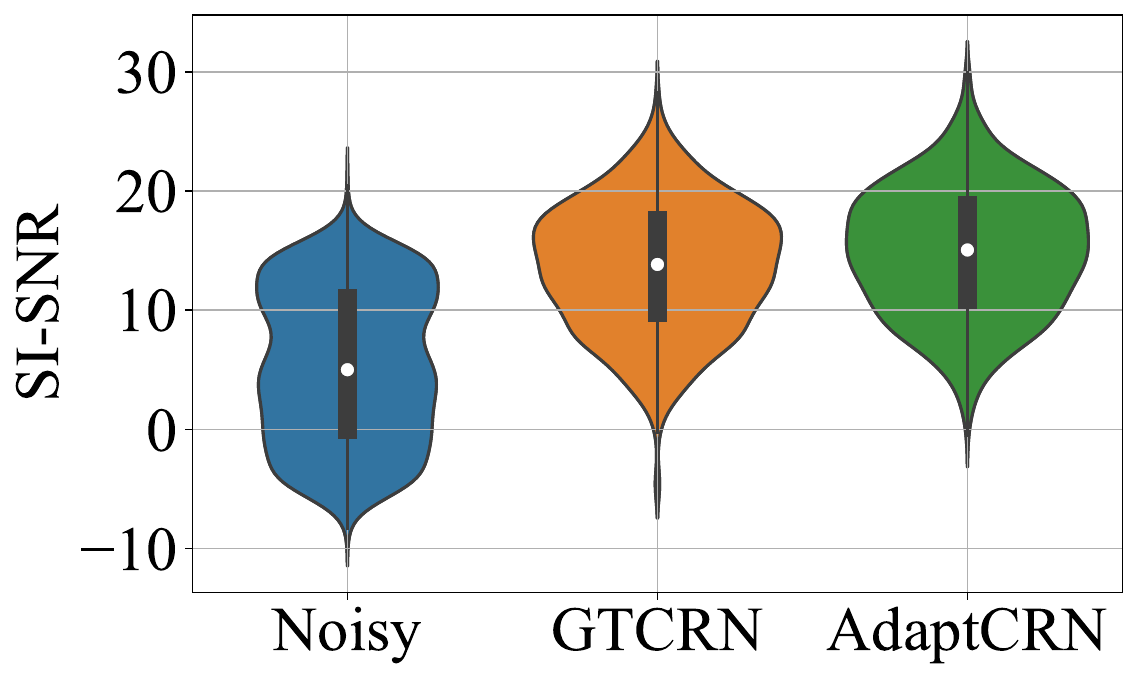}}
    \hfil
    \subfigure[]{\includegraphics[height=1.3in]{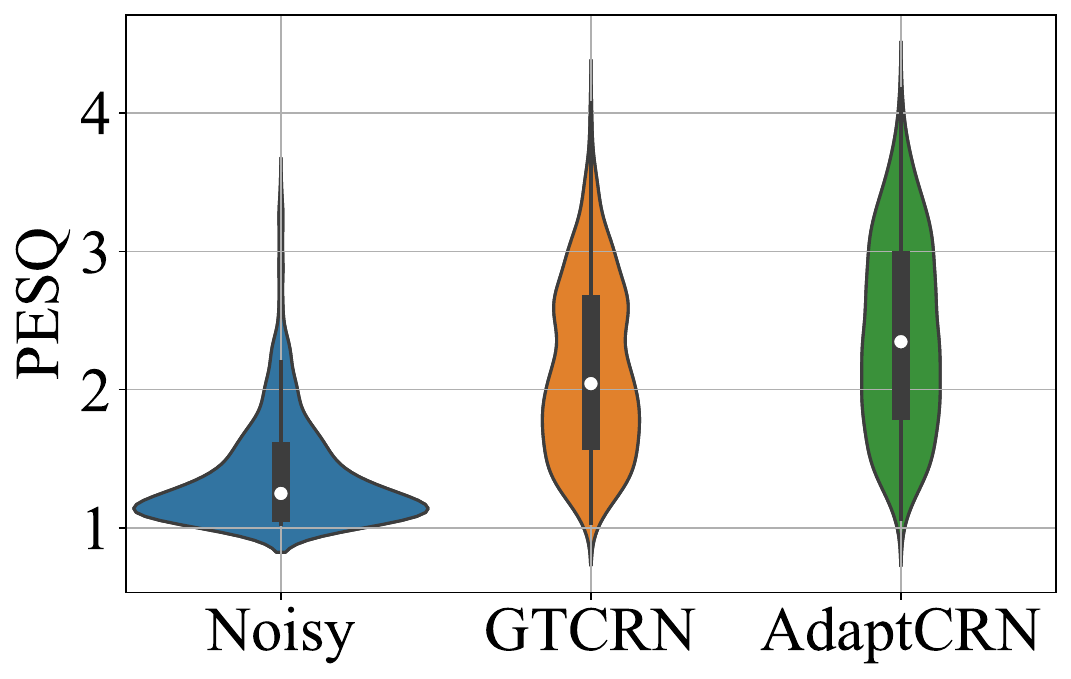}}
    \hfil
    \subfigure[]{\includegraphics[height=1.3in]{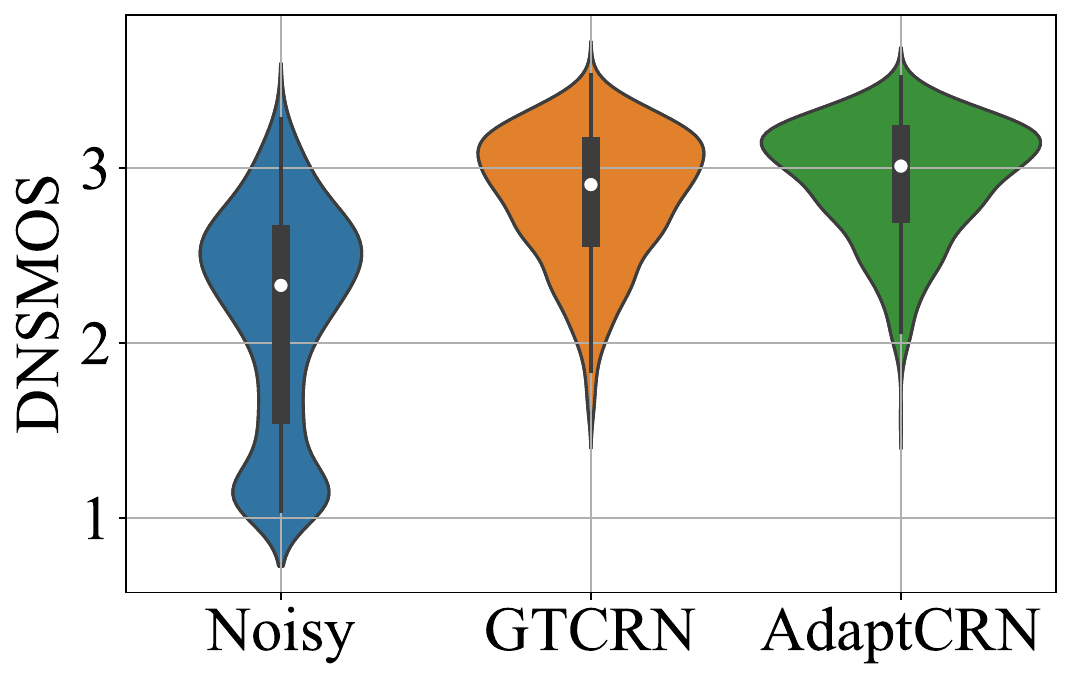}}
    \caption{Violin plots of (a) SI-SNR, (b) PESQ, and (c) DNSMOS-OVRL results for GTCRN and AdaptCRN on the DNS5 test set.}
    \label{fig3}
\end{figure*}

\begin{table}[t]
\renewcommand{\arraystretch}{1.1}
\caption{\label{Table6} Performance of AdaptCRN on Voicebank+DEMAND test set.}
\centering
\begin{tabular}{p{2.2cm}<{\centering}p{1.05cm}<{\centering}p{1.27cm}<{\centering}p{0.90cm}<{\centering}p{0.55cm}<{\centering}p{0.55cm}}
\Xhline{1pt}
\noalign{\vskip 1.5pt}
Model & Para. (K) & MACs (M) & SI-SNR & STOI & PESQ \\
\noalign{\vskip 1.5pt}
\hline
Noisy & - & - & 8.45 & 0.921 & 1.97 \\
DeepFilterNet \cite{schroter2022deepfilternet} & 1780 & 350 & 16.63 & 0.942 & 2.81 \\
GTCRN \cite{rong2024gtcrn} & 24 & 34 & 18.83 & 0.940 & 2.87 \\
ULCNet \cite{shetu2024ultra} & 688 & 98 & 17.20 & - & 2.87 \\
FSPEN \cite{yang2024fspen} & 79 & 89 & - & 0.942 & 2.97 \\
LiSenNet\textsuperscript{$\ast$} \cite{yan2024lisennet} & 37 & 56 & - & 0.937 & 2.95 \\
AdaptCRN & 135 & 41 & 18.82 & 0.940 & 2.98 \\
\Xhline{1pt}
\end{tabular}
\begin{flushleft}
\textsuperscript{$\ast$} The model is trained without PESQ loss. \\
\end{flushleft}
\end{table}

\subsubsection{Ablation study results for AdaptCRN \label{sec6.2.1}}
We evaluate the effectiveness of the proposed structures in AdaptCRN through ablation experiments by selectively removing specific modules. The results, summarized in Table~\ref{Table4}, highlight the pivotal role of adaptive convolution in enhancing model performance. With an increase in computational complexity of approximately 7M MACs, adaptive convolution delivers significant improvements, including nearly 1 dB in SI-SNR, 0.02 in ESTOI, 0.19 in PESQ, and over 0.06 in DNSMOS-OVRL. In the proposed model, we utilize adaptive convolution with $K=8$, as this configuration provides a favorable balance between performance gains and parameter amount. Due to the extremely low parameter count of the basic model, the additional parameters introduced by adaptive convolution remain within an acceptable range for real-time applications.

The results also indicate that joint channel attention contributes positively to the model's performance with minimal increases in both parameters and computational complexity. Replacing the joint multi-layer kernel attention with independent kernel attention for each sub-layer results in negligible performance degradation but introduces over 30K additional parameters and slightly higher MACs, highlighting the efficiency of the joint multi-layer mechanism. Additionally, the dynamic range compression in the spectral compression module demonstrates a positive impact on performance.

Table~\ref{Table4} further presents the model's performance when the activation function between the two PW Conv layers in the adaptive block is removed (DW(8)-PW(8)-skip-PW(8)), where the numbers in parentheses (e.g., 8) denote $K$, the number of candidate kernels. This modification results in some performance degradation but still outperforms a single PW Conv layer with $K=8$ (DW(8)-PW(8)). This indicate that cascading two PW convolution layers with adaptive convolution is not equivalent to a single layer, unlike in the case of vanilla convolution. However, as analyzed in Section~\ref{sec4.3}, this configuration can be reparameterized into a single layer with 64 candidate kernels, showcasing the enhanced nonlinear representational capacity of adaptive convolution. Moreover, directly using 64 independent kernels (DW(8)-PW(64)) achieves better performance compared to the reparameterized result, due to the lower-rank kernel space of the latter. Although this configuration achieves performance comparable to the original AdaptCRN, it requires a larger number of parameters. This result underscores the architectural efficiency of AdaptCRN, which achieves similar performance with fewer redundant parameters.

\subsubsection{Comparison with the baseline models \label{sec6.2.2}}
We evaluate our ultra-lightweight model, AdaptCRN, against DPCRN-light, GTCRN, and LiSenNet on DNS5 test set. As illustrated in Table~\ref{Table5}, AdaptCRN outperforms the baseline models across all metrics. AdaptCRN remains efficient with computational complexity approximately 40M MACs, which is only slightly higher than GTCRN. While it does have more parameters compared to the baselines, it remains within acceptable limits for practical applications. Considering that GTCRN with adaptive convolution has a parameter count and computational complexity similar to AdaptCRN, its results are also presented in Table~\ref{Table5} for comparison. Despite incorporating adaptive convolution, GTCRN performs significantly worse than AdaptCRN, highlighting the effectiveness of the additional structural designs introduced in AdaptCRN. We further present the SI-SNR, PESQ, and DNSMOS-OVRL results of AdaptCRN and the baseline model with comparable computational complexity, GTCRN, using violin plots in Fig.~\ref{fig3}. The results demonstrate that AdaptCRN achieves consistent improvements across all utterances in the test set.

Furthermore, we conduct additional evaluations of AdaptCRN on the Voicebank+DEMAND dataset, using DeepFilterNet, GTCRN, ULCNet, FSPEN, and LiSenNet for comparison. Results for these models are sourced directly from their respective papers. Given the importance of PESQ as a metric in this evaluation, we ensure a fair comparison by using the results for LiSenNet trained without PESQ loss. As shown in Table~\ref{Table6}, the results consistently highlight AdaptCRN's competitive performance alongside lower computational complexity compared to other models.

\begin{figure*}[!t]
    \centering
    \subfigure[]{\includegraphics[width=2.3in]{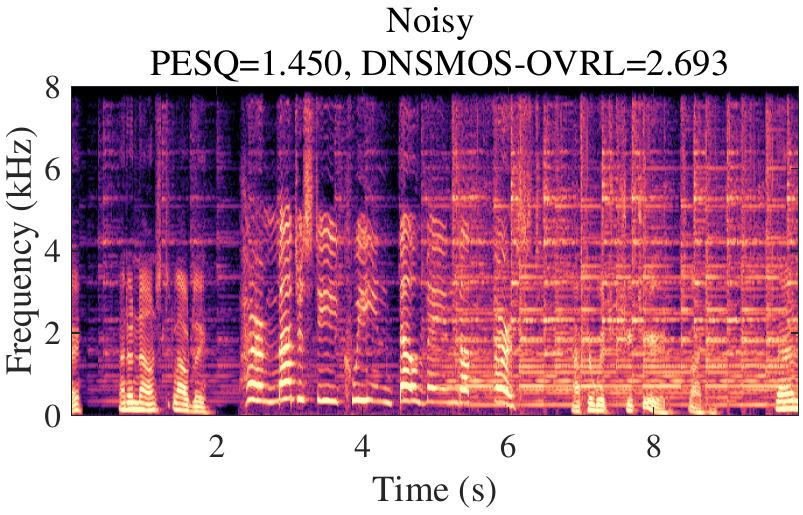}}
    \hfil
    \subfigure[]{\includegraphics[width=2.3in]{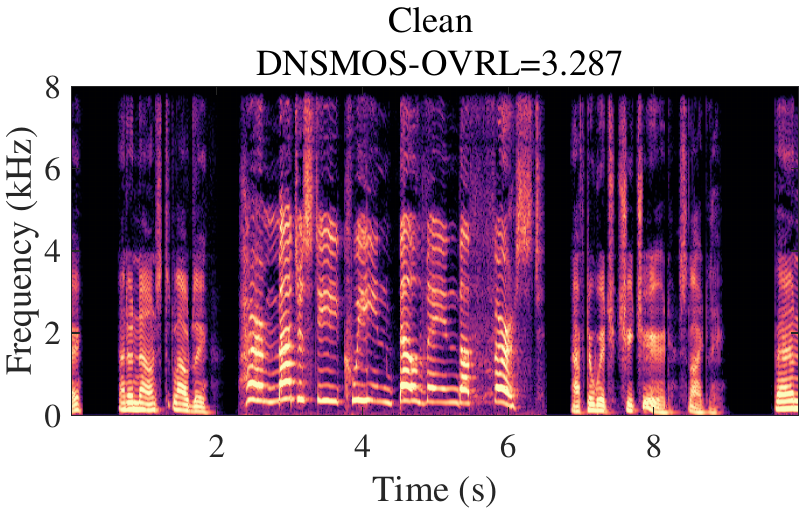}}
    \hfil
    \subfigure[]{\includegraphics[width=2.3in]{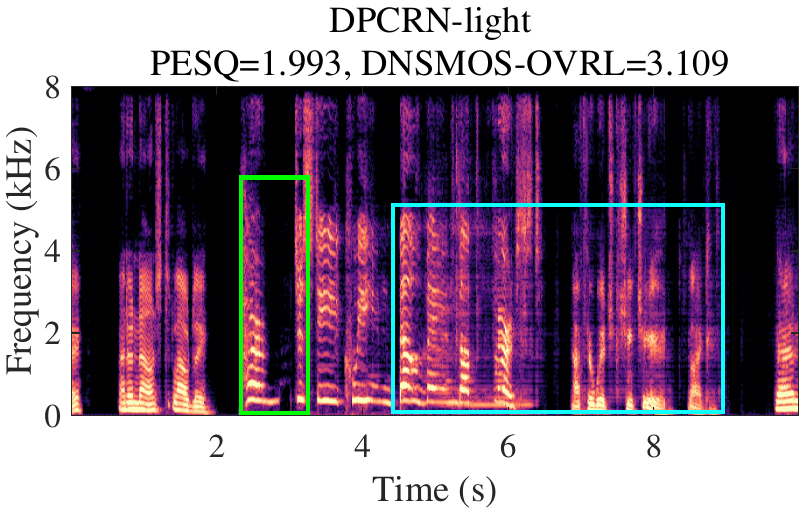}}
    \\
    \subfigure[]{\includegraphics[width=2.3in]{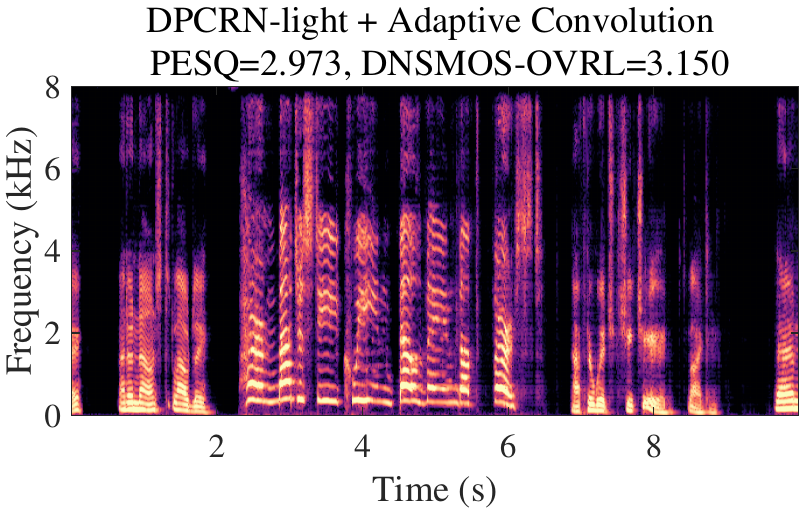}}
    \hfil
    \subfigure[]{\includegraphics[width=2.3in]{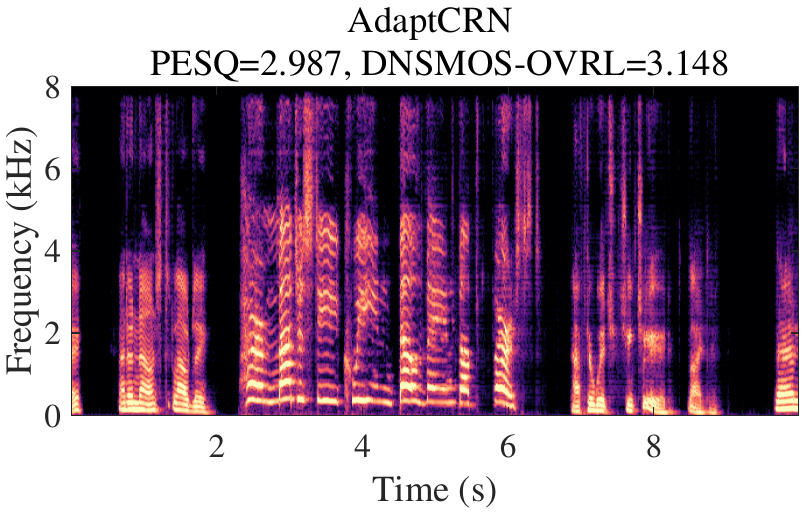}}
    \hfil
    \subfigure[]{\includegraphics[width=2.3in]{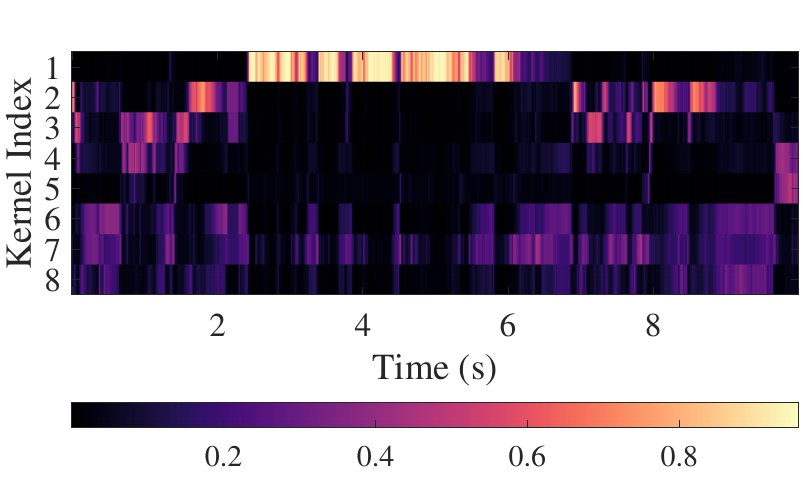}}
    \caption{Example spectrograms from DNS5 test set and kernel attention weight visualization: (a) noisy signal, (b) clean signal, (c) signal enhanced by DPCRN-light, (d) signal enhanced by DPCRN-light with adaptive convolution, (e) signal enhanced by AdaptCRN, (f) visualization of kernel attention weights across frames, derived from the third layer of the decoder in DPCRN-light with adaptive convolution.}
    \label{fig4}
\end{figure*}

\begin{figure}[!t]
    \centering
    \includegraphics[width=3.5in]{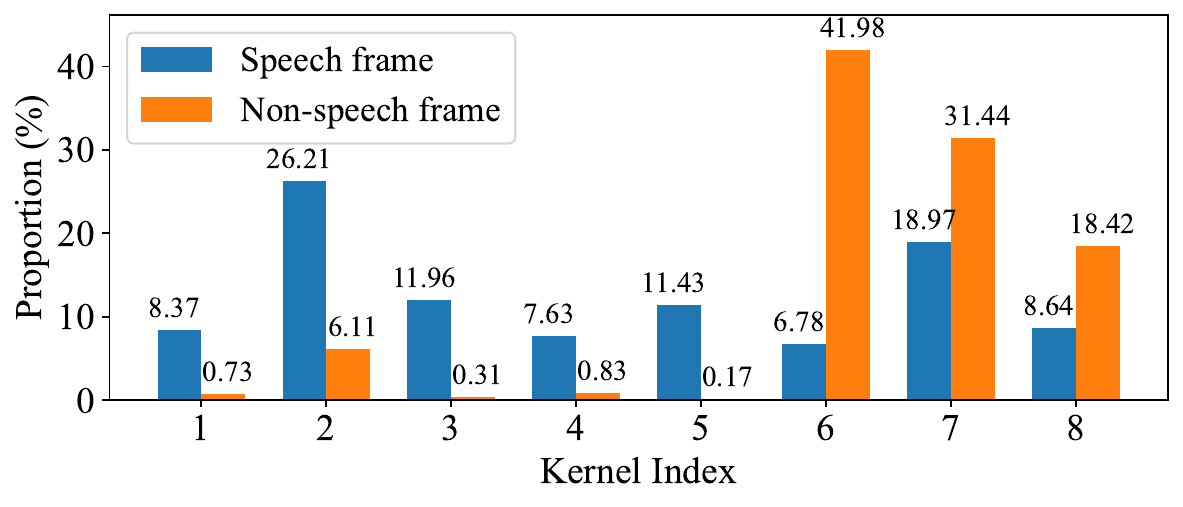}
    \caption{The proportion of frames in which the $k$-th candidate kernel is selected as the dominant kernel across speech or non-speech frames, measured from the third decoder layer of DPCRN-light with adaptive convolution.}
    \label{fig5}
\end{figure}

\subsection{Audio sample visualization \label{sec6.3}}
A typical audio sample from DNS5 test set is presented in Fig.~\ref{fig4}, showcasing noisy and clean spectrograms, enhanced results from DPCRN-light, DPCRN-light integrated with adaptive convolution, and AdaptCRN. This sample poses challenges due to highly non-stationary noise and the presence of two speakers with distinct vocal characteristics, both of which should be preserved in single-channel SE task. The enhanced spectrogram of DPCRN-light exhibits noticeable distortion in one speaker’s voice (highlighted in the green box) and residual noise (highlighted in the blue box), which are effectively mitigated with the incorporation of adaptive convolution. AdaptCRN also achieves superior enhancement quality with ultra-low computational complexity, further highlighting the advantage of the propose methods. Additional audio samples are available at \url{https://github.com/Dahan-Wang/Adaptive-Convolution-for-CNN-based-Speech-Enhancement-Models}.

\subsection{Correlation between kernel allocation and speech spectral features \label{sec6.4}}
We observe that the allocation of candidate kernels in the trained adaptive convolution exhibits a strong correlation with speech spectral features. To illustrate this, the frame-level kernel attention weights are visualized in Fig.~\ref{fig4}(f). The visualization is derived from the third layer of the decoder in DPCRN-light with adaptive convolution, depicting the attention weights of 8 candidate kernels across frames. For clarity, the order of the kernels is manually adjusted during plotting. The results show that the first candidate kernel is predominantly activated during the utterance of the high-pitched speaker, while the segments from the other speaker with a lower pitch are primarily associated with the 2nd to 5th kernels. The 6th to 8th kernels appear to function mainly in pure noise segments. 

We further conduct a quantitative analysis of this relationship on the DNS5 test set. Specifically, we examine the weights assigned to candidate kernels by the kernel attention mechanism of the adaptive convolution layer. For each frame, the candidate kernel with the highest weight is defined as the dominant kernel, as it contributes most strongly to the final convolution kernel. We then compute the proportion of frames in which the $k$-th kernel is selected as the dominant kernel, separately for speech and non-speech frames, with an energy-based voice activity detection (VAD) method used to distinguish between the two categories. The results are shown in Fig.~\ref{fig5}. As observed, in speech frames, all candidate kernels have a noticeable chance of being selected as the dominant kernel, with differences across kernels present but not pronounced. In contrast, in non-speech frames, the 6th-8th kernels are significantly more likely to dominate. This suggests that the last three kernels specialize in processing noise-related features, which is consistent with our visualization results. Notably, these kernels also make substantial contributions in speech frames. Two factors may explain this phenomenon: (1) speech frames still contain noise components, necessitating the involvement of noise-related kernels; and (2) VAD often labels short pauses between words as speech frames, meaning that some ``speech frames'' in practice include non-speech segments, where the last three kernels are more likely to dominate.

These findings indicate that the adaptive convolution assigns appropriate candidate kernels to each frame based on speech characteristics, effectively facilitating feature extraction and reconstruction.

\section{Conclusion \label{sec7}}
In this paper, we propose adaptive convolution, an efficient and versatile convolutional module for CNN-based SE models. It is a frame-wise causal dynamic convolution that adjusts kernels based on spectral characteristics, akin to adaptive filtering. The adaptive convolution kernel is generated by assembling multiple parallel candidate kernels through frame-level attention weights derived from input features. We introduce several lightweight attention mechanisms to guide the kernel aggregation, which can be extended to a multi-head form for joint multi-layer kernel attention and joint channel attention. Moreover, we present AdaptCRN, an ultra-lightweight model that integrates adaptive convolution and several effective strategies. Experimental results and visualizations demonstrate that adaptive convolution can assign appropriate kernels to each frame based on speech characteristics, thereby significantly improving the performance of multiple CNN-based models with a limited increase in computational complexity, especially for lightweight models. The results also show that AdaptCRN achieves competitive performance with several other SOTA lightweight models, even those with significantly higher computational cost, which further underscores the efficiency of adaptive convolution and the effectiveness of the proposed lightweight designs.

\bibliography{IEEEabrv,mybibfile}

\vfill

\end{document}